%% file: main.tex
\documentclass[sigconf, nonacm]{acmart}

\newcommand\vldbdoi{XX.XX/XXX.XX}
\newcommand\vldbpages{XXX-XXX}
\newcommand\vldbvolume{14}
\newcommand\vldbissue{1}
\newcommand\vldbyear{2020}
\newcommand\vldbauthors{\authors}
\newcommand\vldbtitle{\shorttitle} 
\newcommand\vldbavailabilityurl{URL_TO_YOUR_ARTIFACTS}
\newcommand\vldbpagestyle{plain} 

\usepackage{algorithm}

\usepackage{algpseudocode}

\usepackage{booktabs}
\usepackage[table]{xcolor}
\usepackage{makecell}
\usepackage{graphicx}
\usepackage{listings}
\usepackage{xcolor}
\usepackage{listings}
\usepackage[most]{tcolorbox}
\usepackage{verbatim}
\usepackage{bbm}
\usepackage{multirow}

\input{helpers/macros}

\begin{document}
\title{SPA: A SQL-Plan-Aware Reinforcement Learning Framework for Query Rewriting with LLMs}

%%
%% The "author" command and its associated commands are used to define the authors and their affiliations.
\author{Xinyi Huang}
\affiliation{%
  \institution{Simon Fraser University}
  \city{BC}
  \state{Canada}
}
\email{xinyi_huang@sfu.ca}

\author{Zhengjie Miao}
\affiliation{%
  \institution{Simon Fraser University}
  \city{BC}
  \state{Canada}
}
\email{zhengjie@sfu.ca}

% \author{Wang Xiu Ying}
% \author{Zhe Zuo}
% \affiliation{%
%   \institution{East China Normal University}
%   \city{Shanghai}
%   \country{China}
% }
% \email{firstname.lastname@ecnu.edu.cn}

\begin{abstract}

SQL query rewriting is a well-established technique for improving database performance without schema or index changes, yet finding effective rewrites for modern analytical workloads remains difficult: rule-based methods are limited to predefined transformations, while LLM-based approaches often produce rewrites that are semantically valid but compile to equivalent physical plans or degrade runtime performance. 
We present SPA, a SQL-Plan-Aware reinforcement learning framework that trains LLMs to rewrite queries using physical execution feedback. SPA formulates rewriting as a policy optimization problem and extends GRPO with rewards spanning semantic equivalence, textual rewrite distance, physical-plan divergence, and runtime speedup.
To handle reward sparsity across query difficulty, SPA introduces Probability-Gated Adaptive Reward Shaping, a query-level curriculum that unlocks higher-level rewards only once a rollout group achieves sufficient mastery of lower-level objectives, and further improves sample efficiency through on-policy self-improvement by recycling slowdown rewrites from the current policy as targeted training signals. On both IID and OOD workloads, SPA outperforms rule-based and strong LLM baselines in end-to-end runtime, substantially reduces harmful slowdown rewrites, and yields strong tail-latency gains.

\end{abstract}

\maketitle

\pagestyle{\vldbpagestyle}
\begingroup\small\noindent\raggedright\textbf{PVLDB Reference Format:}\\
\vldbauthors. \vldbtitle. PVLDB, \vldbvolume(\vldbissue): \vldbpages, \vldbyear.\\
\href{https://doi.org/\vldbdoi}{doi:\vldbdoi}
\endgroup
\begingroup
\renewcommand\thefootnote{}\footnote{\noindent
This work is licensed under the Creative Commons BY-NC-ND 4.0 International License. Visit \url{https://creativecommons.org/licenses/by-nc-nd/4.0/} to view a copy of this license. For any use beyond those covered by this license, obtain permission by emailing \href{mailto:info@vldb.org}{info@vldb.org}. Copyright is held by the owner/author(s). Publication rights licensed to the VLDB Endowment. \\
\raggedright Proceedings of the VLDB Endowment, Vol. \vldbvolume, No. \vldbissue\ %
ISSN 2150-8097. \\
\href{https://doi.org/\vldbdoi}{doi:\vldbdoi} \\
}\addtocounter{footnote}{-1}\endgroup

% \ifdefempty{\vldbavailabilityurl}{}{
% \vspace{.3cm}
% \begingroup\small\noindent\raggedright\textbf{PVLDB Artifact Availability:}\\
% The source code, data, and/or other artifacts have been made available at \url{\vldbavailabilityurl}.
% \endgroup

\input{sections/intro}

\input{related_work}

\begin{figure}
  \centering
  \includegraphics[width=\columnwidth]{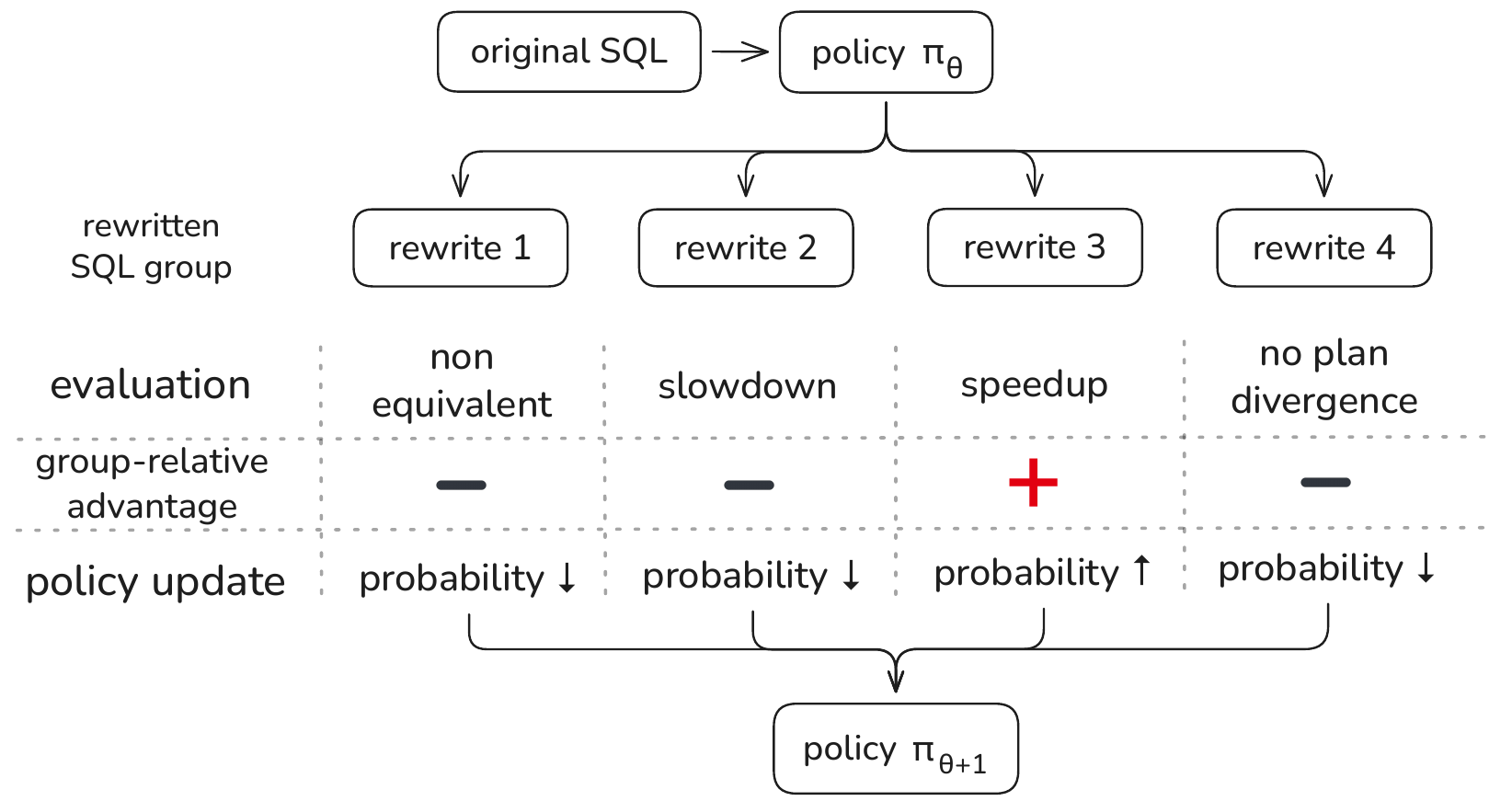}
\caption{Intuition of group-relative policy optimization for SQL rewriting. For the same original SQL query, the policy samples a group of candidate rewrites, and each rewrite is evaluated using database-grounded feedback. Rewrites with positive group-relative advantage are reinforced, while rewrites with negative relative advantage are discouraged.}
  \label{fig:intuitive}
\end{figure}

\begin{figure*}
  \centering
  \includegraphics[width=\linewidth]{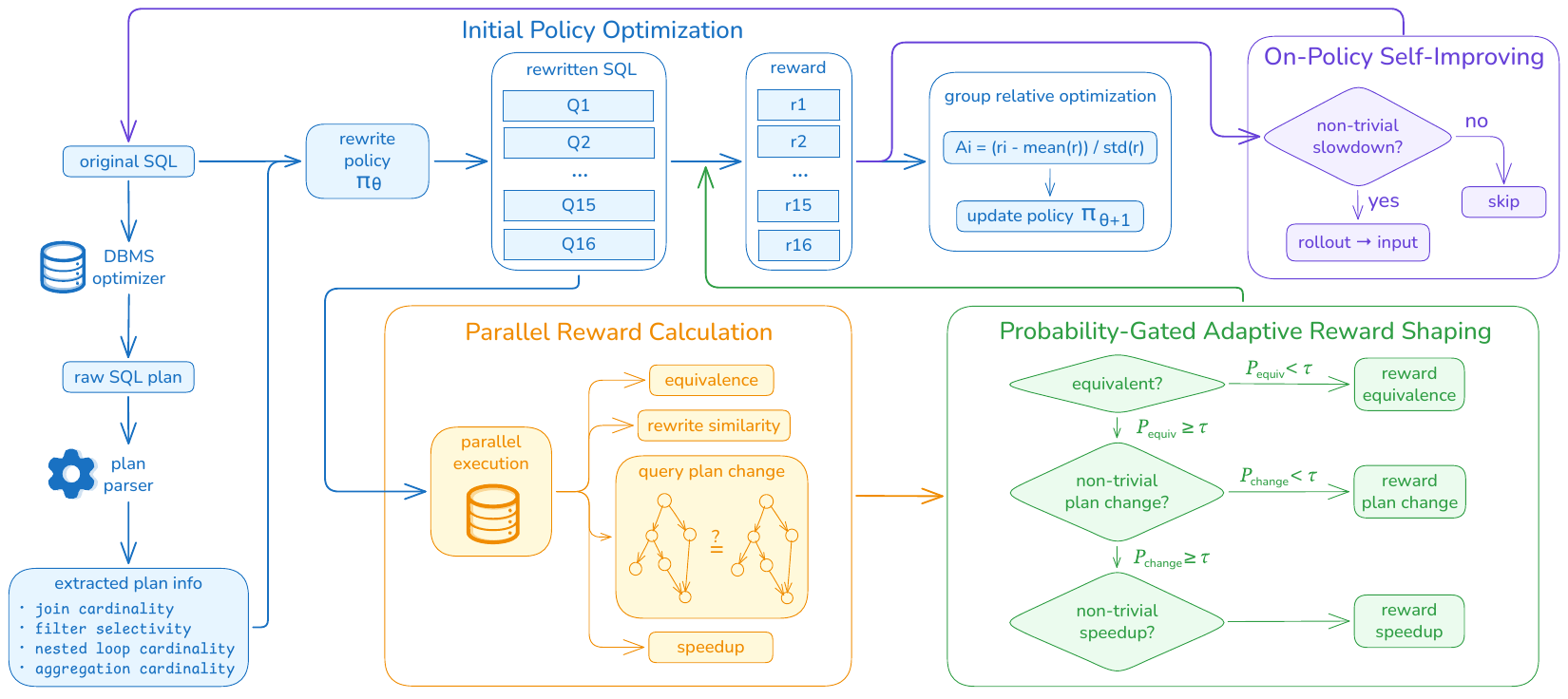}
  \caption{Overview of SPA. SPA trains an SQL rewrite policy using database-grounded feedback from semantic equivalence, physical-plan divergence, and execution latency. The framework combines a GRPO training loop, probability-gated adaptive reward shaping, parallel reward calculation, and on-policy self-improvement.}
  \label{fig:framework}
\end{figure*}

\section{Method}

\subsection{SPA Overview}
\label{sec:overview}

Figure~\ref{fig:framework} presents the training pipeline of SPA. SPA trains an LLM rewrite policy $\pi_\theta$ to transform an original SQL query $x$ into a semantically equivalent rewrite $y$ with improved execution performance. Instead of only imitating existing rewrite examples, SPA directly optimizes the policy using database-grounded feedback from semantic validation, physical-plan behavior, and execution latency.

\paragraph{\textbf{Initial Policy Optimization.}}
SPA builds on Group Relative Policy Optimization (GRPO), which compares multiple sampled outputs for the same input. For each original query $x$, the current policy $\pi_\theta$ samples a group of candidate rewrites $\mathcal{G}=\{y_1,\ldots,y_G\}$. Each candidate rewrite $y_i$ is evaluated by the database-grounded reward pipeline and assigned a scalar reward $r_i$. SPA then computes a group-relative advantage by normalizing rewards within the same rollout group:
\begin{equation}
A_i = \frac{r_i - \mu_{\mathcal{G}}}{\sigma_{\mathcal{G}}+\epsilon},
\end{equation}
where $\mu_{\mathcal{G}}$ and $\sigma_{\mathcal{G}}$ are the mean and standard deviation of rewards in $\mathcal{G}$, and $\epsilon$ is a small constant for numerical stability. The policy is updated to increase the likelihood of candidates with positive advantages and decrease the likelihood of candidates with negative advantages. In practice, this update is performed with a clipped policy-gradient objective, which constrains the change from the old policy $\pi_{\theta_{\mathrm{old}}}$ to the updated policy $\pi_\theta$ for stable training. This group-relative update is well suited to SQL rewriting because all candidates in a rollout group share the same original query and database context, making their execution outcomes directly comparable. Figure~\ref{fig:intuitive} gives an intuitive view of this process.

\paragraph{\textbf{Parallel reward calculation.}}
The reward calculation module evaluates each candidate rewrite against the target DBMS. This step is expensive because it may require query execution, semantic validation, physical-plan extraction, plan comparison, and latency measurement. SPA therefore parallelizes reward calculation across candidates in a rollout group. In addition to measuring speedup, the reward module checks whether the rewrite induces a physical-plan change after the DBMS optimizer processes the query. This is important because two SQL queries may look syntactically different but still compile to the same physical plan, resulting in a trivial rewrite with little optimization value. SPA also uses a textual difference constraint based on Levenshtein distance to discourage reward hacking, where the model returns a syntactically near-identical query or performs only superficial edits while still satisfying validity checks. Together, physical-plan comparison and textual-distance filtering help reduce noisy rewards from trivial rewrites and encourage the policy to search for rewrites that meaningfully affect execution.

\paragraph{\textbf{Probability-gated adaptive reward shaping.}}
To stabilize training, SPA introduces Probability-Gated Adaptive Reward Shaping (PGARS), a group-wise curriculum mechanism for composing the reward. SQL rewrite learning involves multiple objectives with different difficulty levels: the model must first produce syntactically valid SQL, then preserve semantic equivalence, then induce a nontrivial physical-plan change, and finally improve runtime. Rewarding all objectives equally from the beginning can introduce noisy gradients, because early in training many rollouts may be invalid or non-equivalent. PGARS addresses this by activating higher-level reward components only when the rollout group demonstrates sufficient proficiency on prerequisite milestones. For example, speedup reward is activated only after the group shows adequate success in semantic equivalence and plan divergence. This encourages the policy to progress from correctness to execution improvement in a query-adaptive manner.

\paragraph{\textbf{On-policy self-improvement.}}
After the base GRPO training stage, SPA further improves the policy using its own rollout history. The key observation is that slowdown rewrites are not merely failed samples; they can reveal where the current policy misjudges execution behavior. SPA mines slowdown-inducing rewrites that are semantically equivalent but slower than the original query. To avoid revisiting queries that may not be optimizable, SPA further checks the rollout group: a slowdown case is selected only if another candidate rewrite in the same group achieves a nontrivial speedup. Such a group indicates that the query can be optimized, but the current policy has not reliably learned which transformation is beneficial. SPA then blends these selected self-improvement samples with the original training data and continues GRPO training. This on-policy stage focuses learning on high-regret but optimizable cases, helping the model distinguish harmful plan changes from beneficial ones.

Overall, SPA trains a SQL rewrite policy through database-grounded reinforcement learning. The training loop uses GRPO to compare multiple rewrites for the same query, the parallel reward module makes execution-based feedback practical, PGARS stabilizes reward learning through a group-wise curriculum, and on-policy self-improvement reuses the policy's own slowdown cases to improve robustness.

\subsection{Plan Parser}
\label{sec:plan_parser}
To provide the LLM with sufficient information for identifying query bottlenecks before rewriting, we implement a \textit{Plan Parser} that processes PostgreSQL plans generated by \texttt{EXPLAIN (FORMAT JSON)}. The parser extracts key execution signals from the JSON plan and organizes them into concise, readable hints, ensuring that the prompt exposes the relevant physical-plan context needed to guide rewrite decisions.

\paragraph{Cardinality Flow Analysis}
At each plan node, the parser derives a \textit{cardinality delta} from three quantities:
\begin{itemize}
    \item \textbf{Rows In ($R_{in}$):} For leaf scans, the total physical rows read, including those removed by predicates. For internal nodes, the sum of rows emitted by all children.
    \item \textbf{Rows Out ($R_{out}$):} Rows emitted after the operator's logic is applied.
    \item \textbf{Change Ratio:} A label classifying the node as a \textit{reducer} (e.g., ``reduce by 90\%''), an \textit{expander} (e.g., ``expand by 2x''), or a \textit{pass-through}.
\end{itemize}
Pass-through nodes---such as \texttt{Sort} or \texttt{Gather}---that leave cardinality unchanged are pruned to keep the context window compact. Exceptions are made for roots of correlated subplans and high-cost scans, which are always retained.

\paragraph{Semantic Categorization and Grouping}
The parsed nodes are then assigned to five functional categories. Each category pairs cardinality metrics ($R_{in}$, $R_{out}$, Change Ratio) with the operation type, so the LLM can interpret data-flow behavior in terms of concrete relational patterns:
\begin{itemize}
    \item \textbf{Nested Loop:} Captures outer/inner table relationships and per-outer-row counts alongside the resulting cardinality.
    \item \textbf{Join:} Groups Hash and Merge joins by predicate, using $R_{out}$ to reflect join-algorithm efficiency.
    \item \textbf{Join with Filter:} Handles nodes that combine an index-based join with a secondary filter, attributing row reduction to both.
    \item \textbf{Filter:} Covers simple relation scans, mapping predicate selectivity directly to the cardinality delta.
    \item \textbf{Aggregation:} Isolates grouping operations; $R_{out}$ here reflects the cardinality of the result set after grouping.
\end{itemize}
Within each category, nodes are ranked by absolute reduction ($R_{in} - R_{out}$), surfacing the operators responsible for the largest cardinality changes and giving the LLM focused signals about operators that may contribute to execution bottlenecks.

\subsection{Training Data Preparation}

To improve the robustness and generalizability of SPA, we curate a comprehensive training corpus derived from industry-standard schemas.

\paragraph{Workload Generation}

We select TPC-DS~\cite{tpcds} and TPC-H~\cite{tpch}, as they provide complex relational patterns and are widely used industry-standard benchmarks. In addition to using official benchmark toolkits to generate standard analytical workloads, we also employ SQLStorm~\cite{SQLStorm} to synthesize workloads with diverse SQL structures beyond benchmark templates. These generated queries are used for both Supervised Fine-Tuning (SFT) and Reinforcement Learning (RL).

\begin{figure}
  \centering
  \includegraphics[width=0.75\columnwidth]{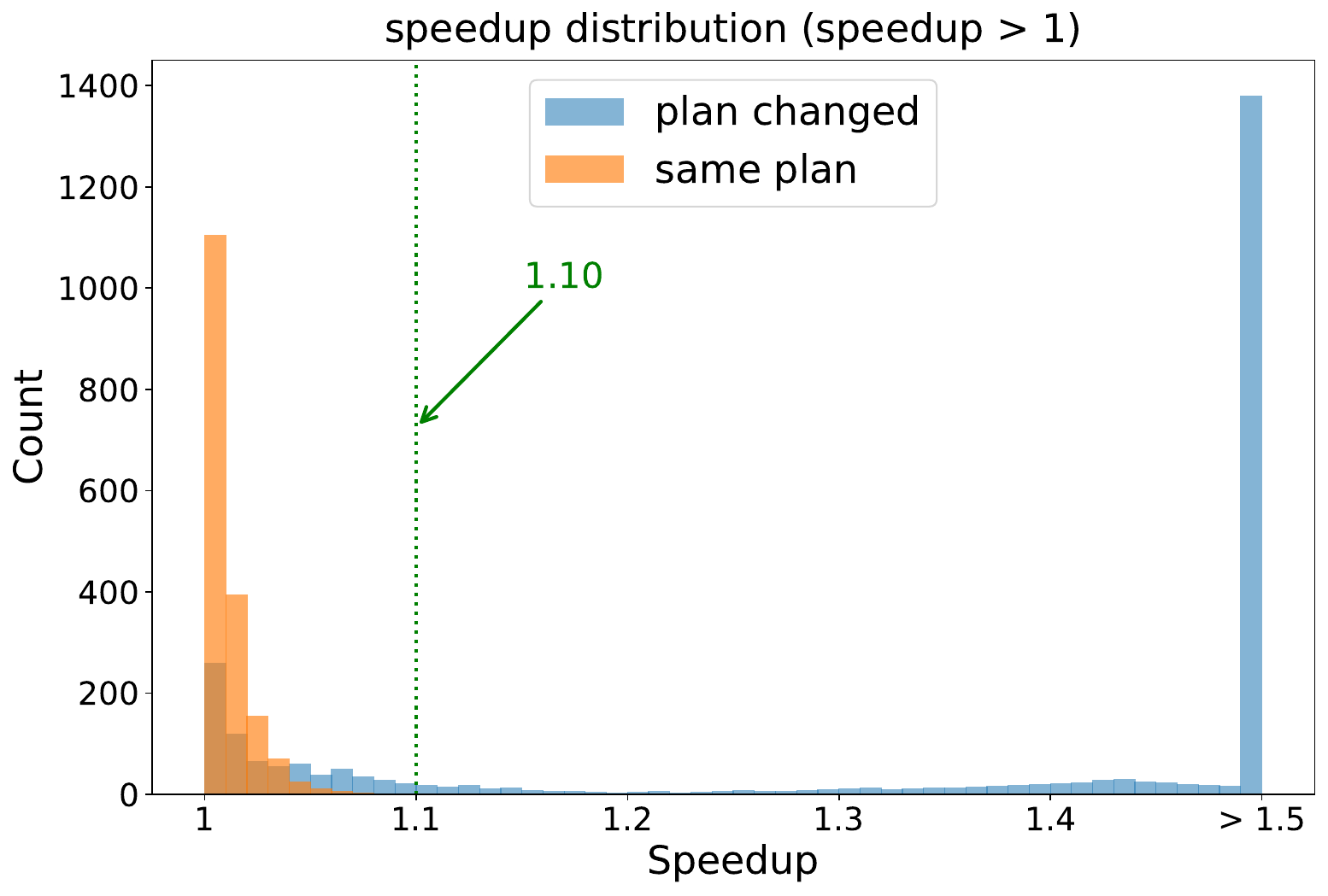}
\caption{Speedup distribution of semantically equivalent query pairs collected during SFT data generation, grouped by whether the physical plan changes.}
  \label{fig:speedup_distribution}
\end{figure}

\paragraph{Rewrite Generation for SFT}
To construct a high-quality dataset for the initial SFT phase, we utilize o3-mini and GPT-5.4 to generate a pool of candidate rewrites. To improve data quality and avoid training on suboptimal or trivial examples, a rewrite is only promoted to the SFT training set if it satisfies two strict criteria:
\begin{enumerate}
    \item \textbf{Semantic Equivalence:} The rewrite must produce a result set identical to the original query.
    \item \textbf{Non-Trivial Performance Gain:} The rewrite must achieve a nontrivial speedup $S > 1.1$.
\end{enumerate}
We choose the threshold $1.1$ based on the empirical speedup distribution shown in Figure~\ref{fig:speedup_distribution}: rewrites with $S > 1.1$ are strongly associated with plan changes, suggesting that they reflect meaningful optimization behavior rather than trivial execution fluctuations.

\subsection{Parallel Reward Calculation}
\label{sec:parallel_reward_calculation}

Execution-based reward evaluation provides high-fidelity feedback for SQL rewriting, but evaluating candidate rewrites sequentially would make GRPO training prohibitively slow. To address this bottleneck, SPA implements a \textit{Parallel Reward Calculation} layer. During each GRPO rollout, the policy generates a group of candidate rewrites, which are dispatched to a pool of isolated database workers. Each worker evaluates one candidate rewrite against the target DBMS and returns the reward signals required for policy optimization. To improve execution throughput during training, we configure the database with \texttt{shared\_buffers} set to 32GB and \texttt{work\_mem} set to 256MB.

As illustrated in Figure~\ref{fig:framework}, each worker computes four reward-related signals: semantic equivalence, textual rewrite distance, physical-plan divergence, and runtime speedup.

% \paragraph{\textbf{Equivalence Verification.}}
% The worker first verifies whether the rewritten query is semantically equivalent to the original query using result-set comparison. This step ensures that the reward does not encourage rewrites that improve latency by changing query semantics. In practice, equivalence verification can be inconclusive when either the original query or the rewritten query reaches the execution timeout. We handle these cases as follows. During training, if original timeout, this group optimization will be skipped since equivalence can not be checked. If the rewritten query times out but the original query finishes, the rewrite receives a negative signal.

\paragraph{\textbf{Equivalence Verification.}}
The worker first verifies whether the rewritten query is semantically equivalent to the original query through result-set comparison. This step prevents the reward from favoring rewrites that reduce latency by changing query semantics. In practice, equivalence verification may be inconclusive when either the original query or the rewritten query reaches the execution timeout. We handle such cases as follows. During training, if the original query times out, the corresponding rollout group is skipped because equivalence cannot be reliably checked. If the rewritten query times out while the original query finishes, the rewrite is treated as unsuccessful and assigned a negative signal.

\paragraph{\textbf{Levenshtein Distance of Rewrite.}}
In addition to semantic validation, SPA measures the textual difference between the original query and the rewritten query using normalized Levenshtein distance. This signal is used to discourage degenerate or lazy rewrites that pass syntactic and semantic checks but make only superficial changes, such as adding redundant parentheses, renaming aliases, or reformatting the SQL text. Let $d_{\mathrm{lev}}(x,y)$ denote the Levenshtein edit distance between the original query $x$ and the rewrite $y$. We normalize it by the maximum query length:
\[
D_{\mathrm{lev}}(x,y) = \frac{d_{\mathrm{lev}}(x,y)}{\max(|x|,|y|)}.
\]
A rewrite is considered textually nontrivial only when $D_{\mathrm{lev}}(x,y)$ exceeds a threshold $\lambda$. In our experiments, we set $\lambda=0.01$. This distance constraint helps prevent reward hacking where the model repeatedly returns near-identical SQL strings to obtain correctness rewards without exploring useful rewrite transformations.

% \begin{algorithm}
% \caption{Recursive Plan Tree Comparison}
% \label{alg:plan_comparison}
% \begin{algorithmic}[1]
% \Function{IsDifferent}{Node $n_1$, Node $n_2$}
%     \State \Comment{Check operator type and metadata}
%     \If{$type(n_1) \neq type(n_2)$ \textbf{or} $props(n_1) \neq props(n_2)$}
%         \State \Return \textbf{true}
%     \EndIf

%     \State \Comment{Check child count parity}
%     \If{$len(children(n_1)) \neq len(children(n_2))$}
%         \State \Return \textbf{true}
%     \EndIf

%     \State \Comment{Recursive depth-first comparison}
%     \For{$i \gets 0$ \textbf{to} $len(children(n_1)) - 1$}
%         \If{\Call{IsDifferent}{$children(n_1)[i], children(n_2)[i]$}}
%             \State \Return \textbf{true}
%         \EndIf
%     \EndFor

%     \State \Return \textbf{false} \Comment{Plans are structurally identical}
% \EndFunction
% \end{algorithmic}
% \end{algorithm}

\paragraph{\textbf{Query Plan Divergence Analysis.}}
SPA further checks whether the rewrite induces a different physical execution plan after DBMS optimization. This step is important because two SQL queries may look different at the text level but still be normalized by the optimizer into the same physical plan. Such rewrites are unlikely to yield nontrivial performance improvement and can introduce noisy reward signals if treated as meaningful optimizations.

To detect plan divergence, each worker obtains the physical plans of both the original and rewritten queries using \texttt{EXPLAIN}. We then compare the two plans structurally. Each plan is treated as an operator tree whose nodes correspond to physical operators such as \texttt{SeqScan}, \texttt{IndexScan}, \texttt{HashJoin}, and \texttt{NestedLoop}. SPA recursively compares the two trees by checking operator types, critical operator properties, and child structure. Critical properties include join keys, filter predicates, scan relations, and aggregation keys. A rewrite is considered plan-divergent if any corresponding node differs in operator type, relevant properties, or child structure. 

% Algorithm~\ref{alg:plan_comparison} summarizes this recursive comparison procedure.

At the same time, SPA only proceeds to runtime speedup measurement when a nontrivial plan divergence is detected. This avoids spending expensive execution budget on rewrites that are likely to compile to the same physical plan and therefore provide negligible optimization benefit.

\paragraph{\textbf{Speedup Measurement.}}
For rewrites that pass semantic verification and induce a nontrivial plan change, the worker measures execution latency and computes the speedup ratio:
\[
S = \frac{\mathrm{Latency}_{\mathrm{original}}}{\mathrm{Latency}_{\mathrm{rewrite}}}.
\]
A value of $S>1$ indicates that the rewrite is faster than the original query, while $S<1$ indicates a slowdown. To reduce sensitivity to small runtime fluctuations, SPA uses a nontrivial speedup threshold $\delta$ and maps the raw speedup rate into a bounded reward:
\[
R_s(S) =
\begin{cases}
\tanh\left(\ln \frac{S}{1-\delta}\right), & S < 1-\delta, \\
0, & 1-\delta \le S \le 1+\delta, \\
\tanh\left(\ln \frac{S}{1+\delta}\right), & S > 1+\delta.
\end{cases}
\]
We set $\delta=0.1$ in our experiments. Thus, small latency changes within $\pm 10\%$ are treated as neutral, while clear speedups and slowdowns receive positive and negative rewards, respectively.

\subsection{Probability-Gated Adaptive Reward Shaping}
\label{sec:pga}

Assessing the optimization difficulty of a SQL query through static inspection alone is unreliable. A syntactically simple query may admit only a narrow space of valid rewrites, while a structurally complex query might offer numerous opportunities for profitable transformation. To navigate this variability, we propose \textbf{Probability-Gated Adaptive Reward Shaping (PGARS)}, a mechanism that functions as a query-level curriculum. It dynamically shifts the optimization target according to the policy's observed mastery within each rollout group, ensuring stable convergence across diverse query complexities.

Rather than rewarding all optimization objectives simultaneously, which often leads to unstable gradients and sub-optimal local minima in early training, PGARS decomposes the task into four hierarchical milestones: \textit{Syntax Validity}, \textit{Semantic Equivalence}, \textit{Plan Divergence}, and \textit{Runtime Speedup}. Under this gated scheme, a reward for a higher-level objective is unlocked only when the rollout group exhibits consistent success in all preceding milestones. This ensures the model does not chase performance gains through invalid or non-equivalent rewrites.

Formally, given an input query $x$, the Group Relative Policy Optimization (GRPO) framework generates a rollout group $\mathcal{G} = \{y_1, y_2, \dots, y_G\}$, where $G$ is the number of candidate rewrites and $y_i$ denotes the $i$-th candidate. For each intermediate milestone $k \in \{\text{syn, eq, plan}\}$, we define a binary indicator $\mathbbm{1}_{k,i}$ to signify whether rollout $i$ satisfies that criterion. The group-level proficiency for milestone $k$ is then determined by the success count $c_k = \sum_{i=1}^{G} \mathbbm{1}_{k,i}$. A stage is considered mastered once the group's empirical success rate crosses a predefined gating threshold $\tau \in (0,1]$, expressed by the gate function:
\begin{equation}
\Gamma_k = \mathbbm{1}\left(\frac{c_k}{G} \ge \tau\right)
\end{equation}

The cumulative reward $r_i$ for an individual rollout $y_i$ is formulated as a conditional sum of milestone rewards $R_{\text{syn}}, R_{\text{eq}}, R_{\text{plan}}$ and the measured speedup rate $R_s(S)$. By applying the gates $\Gamma$ as coefficients, we ensure that latency improvements only contribute to the gradient once the model demonstrates stable correctness and plan-level competence:
\begin{equation}
\label{eq:gated_reward}
r_i = \mathbbm{1}_{\text{syn},i} R_{\text{syn}} + \Gamma_{\text{syn}} (\mathbbm{1}_{\text{eq},i} R_{\text{eq}}) + \Gamma_{\text{syn}} \Gamma_{\text{eq}} (\mathbbm{1}_{\text{plan},i} R_{\text{plan}}) + \Gamma_{\text{syn}} \Gamma_{\text{eq}} \Gamma_{\text{plan}} R_s(S)
\end{equation}

The gating threshold $\tau$ controls how conservatively PGARS advances through the reward hierarchy:
\begin{itemize}
    \item \textbf{High thresholds ($\tau \to 1$):} A gate opens only when most rollouts satisfy the corresponding milestone, making reward progression more conservative.
    \item \textbf{Low thresholds ($\tau \to 0$):} A gate opens with fewer successful rollouts, making reward progression more aggressive.
\end{itemize}

Table~\ref{tab:pgars_example} illustrates PGARS with rollout group size $G=8$ and gating threshold $\tau=25\%$. Under this setting, the next gate opens only if at least two rollouts satisfy the current milestone. Since only one rollout produces a semantically equivalent rewrite, the plan-divergence and runtime-speedup rewards are not activated. This prevents the policy from prematurely optimizing for speedup before it can reliably generate equivalent rewrites.

\begin{table}[t]
\centering
\caption{Example of PGARS with group size $G=8$ and threshold $\tau=25\%$. A gate becomes active when at least 2 rollouts satisfy the corresponding milestone.}
\label{tab:pgars_example}
\resizebox{0.9\columnwidth}{!}{\begin{tabular}{ccccc}
\toprule
\textbf{Rollout} & \textbf{Syntax Valid} & \textbf{Semantic Equiv.} & \textbf{Plan Divergence} & \textbf{Speedup} \\
\midrule
$y_1$ & \checkmark & \checkmark & -- & -- \\
$y_2$ & \checkmark & $\times$ & -- & -- \\
$y_3$ & \checkmark & $\times$ & -- & -- \\
$y_4$ & \checkmark & $\times$ & -- & -- \\
$y_5$ & \checkmark & $\times$ & -- & -- \\
$y_6$ & \checkmark & $\times$ & -- & -- \\
$y_7$ & $\times$ & $\times$ & -- & -- \\
$y_8$ & $\times$ & $\times$ & -- & -- \\
\midrule
\textbf{Threshold Check} & $6/8 > \tau$ & $1/8 < \tau$ & -- & -- \\
\textbf{Next Gate Status} & active & inactive & -- & -- \\
\bottomrule
\end{tabular}}
\end{table}

\begin{figure}[htbp]
    \centering
    \includegraphics[width=0.45\textwidth]{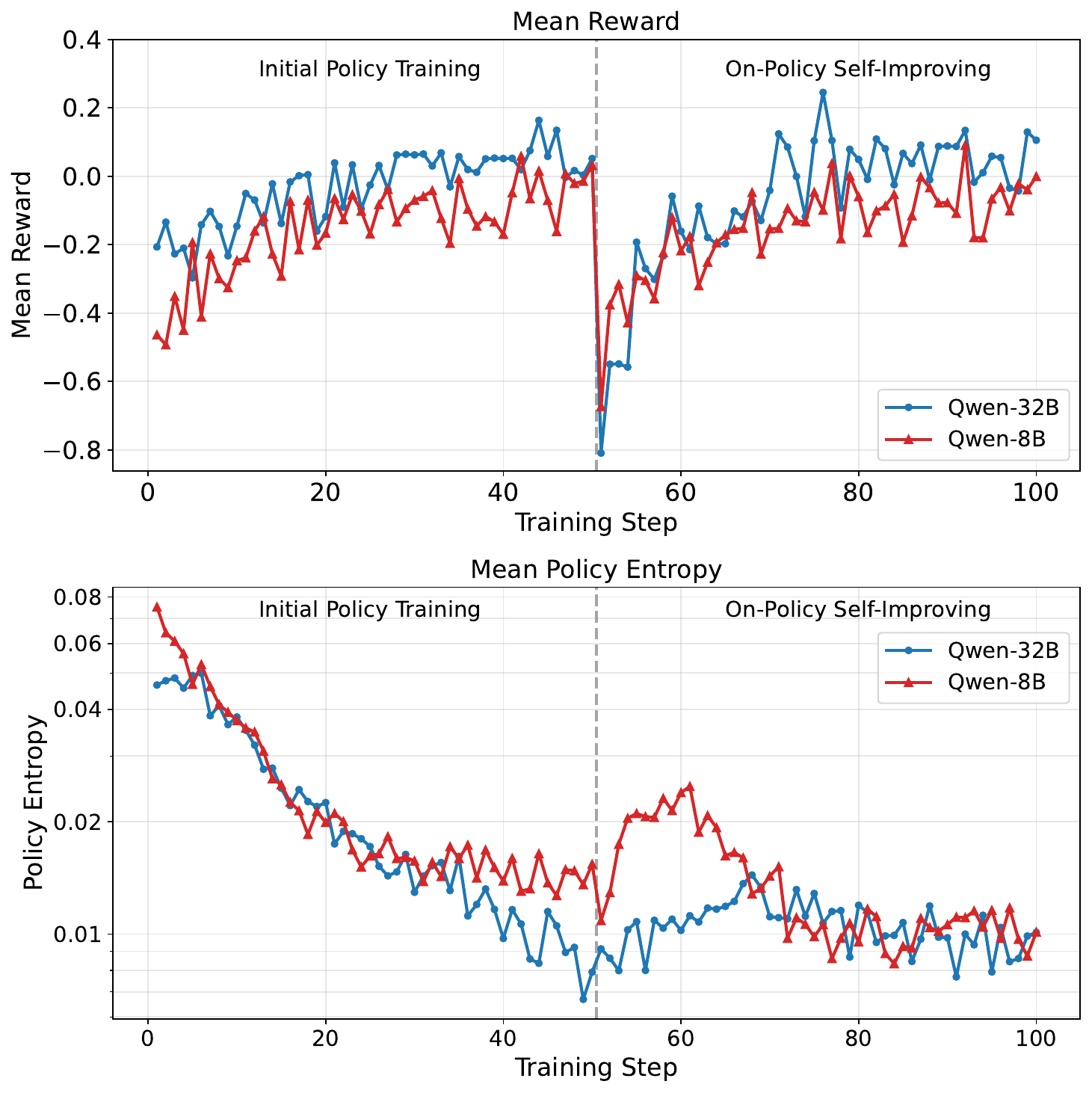} 
    \caption{Mean reward and policy entropy in initial policy optimization and on-policy self-improvement stage.}
    \label{fig:self-improvement}
\end{figure}

\subsection{On-Policy Self-Improvement}
\label{sec:self_improving}

% The empirical analysis in Table~\ref{tab:trivial_rewrite_trap} reveals a persistent challenge in LLM-based query optimization, which we refer to as the \emph{slowdown trap}. A slowdown rewrite is a semantically equivalent rewrite that changes the query but executes slower than the original query. Across DSB, TPC-DS, and TPC-H, such rewrites account for 7.18\%, 23.31\%, and 34.40\% of cases, respectively, resulting in an aggregate slowdown rate of 21.97\%. These results suggest that execution-aware rewriting remains challenging even for strong LLM-based rewriters: a rewrite can be syntactically plausible and semantically valid, yet still induce an unfavorable physical plan.

Unlike general reasoning tasks where the model output is an answer in a different space from the input, SQL rewriting is closed over the query space: the output of a rewrite is itself an executable SQL query and can therefore be used as the input to another rewrite. This property enables SPA to construct recovery workloads from the policy's own execution regressions. Specifically, a slowdown rewrite that remains semantically equivalent to the original query is not discarded as a failed output; instead, it becomes a valid rewriting input whose physical plan has been pushed toward an unfavorable region. Training on such queries encourages the policy to recover from harmful plan changes that it previously introduced.

\paragraph{Slowdown candidate mining.}
To address this issue, SPA introduces an on-policy self-improvement phase after the initial base training stage. The goal of this phase is to reuse the policy's own failed rollouts as targeted training signals. During base training, SPA stores rollout records containing the original query, the generated rewrite, semantic validation result, physical plan, and measured execution latency. For each original query $q$, GRPO samples a group of candidate rewrites $\{\hat{q}_1,\ldots,\hat{q}_G\}$, which allows SPA to compare different execution outcomes for the same query.

We first identify slowdown candidates from the rollout history. A generated rewrite $\hat{q}_i$ is considered a slowdown candidate if it is semantically equivalent to the original query $q$ but has higher measured execution latency:
\[
C_{\mathrm{slow}}(q,\hat{q}_i) = \mathbbm{1}\left[\mathrm{Equiv}(q,\hat{q}_i)=1 \land T(\hat{q}_i)>T(q)\right],
\]
where $T(\cdot)$ denotes measured execution latency and $\mathrm{Equiv}(\cdot)$ denotes semantic-equivalence validation.

\paragraph{\textbf{Group-level optimizability check.}}
Not every slowdown candidate provides useful signal for further training. If all rollouts for a query fail to improve execution, then the query may be difficult to optimize under the current rewrite space, and repeatedly revisiting it may provide limited benefit. Therefore, SPA further checks whether the original query is \emph{optimizable} according to the group-level rollout history.

Specifically, a query $q$ is considered optimizable if at least one other rewrite in the same rollout group achieves a nontrivial speedup:
\[
C_{\mathrm{opt}}(q) = \mathbbm{1}\left[\exists j \in \{1,\ldots,G\}: \mathrm{Equiv}(q,\hat{q}_j)=1 \land \Delta(q,\hat{q}_j)>\epsilon\right],
\]
where $\Delta(q,\hat{q}_j)$ denotes the speedup of rewrite $\hat{q}_j$ over the original query and $\epsilon$ is the threshold for nontrivial improvement.

The final self-improvement set is defined as:
\[
\mathcal{D}_{\mathrm{self}} = \{(q,\hat{q}_i) \mid C_{\mathrm{slow}}(q,\hat{q}_i) \land C_{\mathrm{opt}}(q)\}.
\]
This filtering focuses self-improvement on cases where the policy has observed both harmful and beneficial execution outcomes for the same original query. Such cases are especially valuable because they expose fine-grained distinctions between rewrites that induce unfavorable plans and rewrites that produce meaningful speedups.

\paragraph{\textbf{Data blending.}}
To avoid overfitting to failure cases and to preserve the rewrite ability learned during base training, we blend the selected self-improvement samples with the original base training data using a 1:1 ratio. 
% This mixture maintains exposure to general rewrite patterns while emphasizing difficult cases where the policy has already observed both harmful and beneficial execution outcomes.

\paragraph{\textbf{self-improvement training.}}
The self-improvement phase is \emph{on-policy} because the additional training instances are derived from rollouts generated by the current policy itself. Instead of relying on an external static corpus, SPA uses the policy's own execution outcomes to identify where it remains unreliable. As the policy evolves, the collected slowdown cases reflect the current model's mistakes, allowing the training distribution to adapt toward high-regret but optimizable regions of the rewrite space.

For each selected pair $(q,\hat{q}_i) \in \mathcal{D}_{\mathrm{self}}$, SPA revisits the original query $q$ during the self-improvement phase. The slowdown rewrite $\hat{q}_i$ serves as execution-grounded evidence that the current policy can produce harmful plan changes for this query. The model is then optimized on the blended training distribution using the same database-grounded reward function as in base training, including semantic correctness, physical-plan behavior, and execution-latency improvement.

Overall, the on-policy self-improvement phase turns slowdown rewrites from failed outputs into actionable training signals. By revisiting slowdown cases only when the rollout group demonstrates that improvement is possible, SPA focuses training on queries where better execution behavior is attainable and where the policy can learn to distinguish beneficial plan changes from harmful ones.

As shown in Figure~\ref{fig:self-improvement}, after training shifts to the slowdown-rollout workload, the mean reward drops sharply and requires a similar number of steps as the base training stage to recover. The policy entropy also increases, indicating that the model has not yet mastered the slowdown workload.

\begin{table}[t]
\centering
\small
\caption{Training data size used in different stages of SPA. Slowdown rewritten queries are collected from rewritten queries generated during initial policy optimization stage.}
\label{tab:training_data_size}
\begin{tabular}{ll}
\toprule
\textbf{Stage} & \textbf{Data Size} \\
\midrule
SFT initialization & 5k rewrite pairs \\
\midrule
Initial Policy Optimization & 1.6k original queries \\
\midrule
On-policy self-improvement 
& \makecell[c]{0.8k original queries +\\0.8k slowdown rewritten queries} \\
\bottomrule
\end{tabular}
\end{table}

\input{tables/main_res}

\input{tables/ablation}

\section{Experiment}

\subsection{Experimental Setup}

\subsubsection{Training Setup}

We train SPA with LoRA~\cite{lora}, using rank $r=32$, learning rate $4 \times 10^{-5}$, SQL execution timeout of $120$s, and rollout group size $G=16$. Training is conducted through an API-based training infrastructure. Database-side parallel reward evaluation is performed on a single Ubuntu machine with 128 GiB of memory. We configure PostgreSQL with \textit{shared\_buffers}=32GB and \textit{work\_mem}=256MB to support concurrent query execution. Table~\ref{tab:training_data_size} summarizes the training data size used in each stage.

\subsubsection{Evaluation Setup}

Evaluations are conducted on a PostgreSQL 16.14 server running Ubuntu 24.04, equipped with 8 virtual CPUs and 30\,GiB of RAM. The CPU is an Intel(R) Xeon(R) Platinum 8175M CPU @ 2.50GHz, with 1 socket, 4 cores per socket, and 2 threads per core.

Semantic equivalence is verified through execution-based result comparison. For each query, we first perform one warm-up execution and then report the average runtime over three subsequent executions. For all evaluated models, we use the same retry-with-reflection mechanism during SQL generation. Each generated rewrite is first checked for executability using \texttt{EXPLAIN}. If the rewritten SQL fails this check, the corresponding error message is fed back to the model for another generation attempt. We allow at most three retry attempts per query. To ensure a fair comparison, all LLM baselines are given the same prompt with parsed query plan information, generated by the plan parser described in Section~\ref{sec:plan_parser}.

% If result-set comparison is inconclusive due to timeout, we attempt formal equivalence verification using Parseval~\cite{parseval,qed,sqlsolver}. For unsupported cases, we use GPT-5.4 as a fallback equivalence evaluator.

\subsubsection{Datasets}

To evaluate the performance and robustness of SPA, we partition our benchmarks into In-Distribution (IID) workloads used during training and Out-of-Distribution (OOD) workloads used strictly for zero-shot evaluation. Depending on the benchmark, we either use its official query set directly or generate one concrete SQL query from each provided query template. For toolkit-generated workloads, we use different \texttt{stream\_id}s for training and evaluation queries to avoid overlap.

\paragraph{In-Distribution (IID) Workloads}
The training corpus comprises two widely recognized benchmarks representing standard analytical and relational workloads:
\begin{itemize}
    \item TPC-H~\cite{tpch}. TPC-H is a decision support benchmark, which illustrates decision support systems that examine large volumes of data, execute queries with a high degree of complexity, and give answers to critical business questions.
    \item TPC-DS~\cite{tpcds}. We generate queries using the official toolkits for these industry-standard decision-support benchmarks. They provide a range of complexities, from simple aggregations to multi-way joins across large-scale star and snowflake schemas.
    % \item \textbf{IMDB-JOB (Join Order Benchmark)~\cite{job}.} This dataset consists of queries against the real-world IMDB database. Unlike the synthetic TPC schemas, JOB focuses on complex join predicates and varying selectivity, representing the challenges of real-world data distributions.
\end{itemize}

\paragraph{Out-of-Distribution (OOD) Workloads}
To test whether the model has learned general relational optimization principles rather than simply memorizing training schemas, we evaluate on two distinct OOD workloads:
\begin{itemize}
    \item StackOverflow~\cite{bao}. This dataset represents a significant domain shift, featuring real-world queries from the Stack Overflow data dump. 
    % These queries are typically written by humans with diverse coding styles and focus on web-application patterns rather than analytical reporting.
    \item DSB~\cite{dsb}. Based on a modified TPC-DS schema, DSB introduces highly skewed data distributions and more complex query templates.
\end{itemize}

\subsubsection{Baselines}

To comprehensively evaluate the effectiveness of SPA, we compare our framework against both general-purpose LLMs, base model after supervised fine-tuning (SFT), and strong database-oriented optimization baselines.

\paragraph{\textbf{LLMs}} 
We compare SPA with three models: \texttt{GPT-5.4}~\cite{openai_gpt54_thinking}, \texttt{GPT-4o}~\cite{gpt-4o} and \texttt{Gemini-2.5-Pro}~\cite{gemini-2.5}. We also compare against \texttt{Qwen3-8B} and \texttt{Qwen3-32B}~\cite{qwen3}, which serve as the SPA's backbone models.

\paragraph{\textbf{LearnedRewrite}} 
This system \cite{learnedrewrite} treats query rewriting as a search problem within a policy tree. It employs Monte Carlo Tree Search (MCTS) to navigate the exponential space of rewrite rule sequences. By utilizing a learning-based estimator to predict performance gains, it attempts to find the optimal rule ordering that traditional heuristic optimizers often miss.

\paragraph{\textbf{R-Bot}}
R-Bot~\cite{rbot} is an evidence-guided LLM-based SQL query rewrite system designed to improve query execution efficiency while preserving query semantics. It prepares rewrite evidence from multiple sources, retrieves relevant evidence through a hybrid structure-semantics retrieval method, and uses an LLM to perform step-by-step rule selection and arrangement with self-reflection. Unlike purely heuristic rewrite systems, R-Bot combines retrieved rewrite evidence with LLM reasoning to produce optimized SQL queries. We use \texttt{Gemini-2.5-Pro} as its backbone model.

We exclude E3-Rewrite because it is not publicly available, depends on proprietary LLM-based equivalence judgments, and optimizes estimator-based costs rather than execution-grounded rewards.

% \paragraph{\textbf{LLM-R2}} 
% LLM-R2~\cite{llmr2} is an LLM-enhanced rule-based rewrite system. It addresses the scalability issues of traditional rule enumeration by using an LLM to recommend specific rewrite rules and their optimal application sequences. This approach aims to bridge the gap between flexible LLM reasoning and the formal equivalence guarantees of traditional rule-based systems.

% \subsubsection{Hyperparameters}
% The efficiency and performance of our framework is impacted by several parameters: group size and SQL execution timeout of the GRPO. A detailed analysis of these parameters is provided in Section~\ref{sec:recipe}.

\subsection{Main Results}
\label{sec:main_results}

Table~\ref{tab:main_results} reports the performance of different SQL rewrite methods across four benchmarks. The first two benchmarks, TPC-H and TPC-DS, are IID settings, while DSB and SO are OOD settings. 

Overall, SPA achieves strong and consistent performance across both IID and OOD workloads, especially on tail-latency metrics. On the IID benchmarks, SPA-32B obtains the best or near-best runtime performance. On TPC-H, it achieves the lowest mean runtime of 32.73s, reducing the original mean runtime from 57.00s, and also obtains the best P50, P75, P90, and P95 runtimes among all methods. On TPC-DS, SPA-32B achieves the best P50 and P75 runtimes, while GPT-5.4 obtains the lowest mean runtime and best P90/P95 latency. Compared with the original queries, SPA-32B reduces the mean runtime from 30.92s to 19.92s and improves P95 runtime from 300.00s to 72.92s.

On the OOD workloads, SPA generalizes well and often outperforms strong proprietary models. On DSB, SPA-32B achieves the best mean runtime of 6.96s, substantially lower than the original runtime of 21.23s, and also obtains the best P90 and P95 runtimes. On StackOverflow, SPA-32B maintains a 100\% success rate and 100\% equivalence rate, while achieving competitive runtime performance compared with GPT-5.4 and Gemini-2.5-Pro.

These results show that SPA learns rewrite strategies that transfer to OOD workloads, improving both mean runtime and tail latency while maintaining high success and equivalence rates. In particular, SPA-32B provides the most consistent gains across datasets, suggesting that scaling the policy model further improves rewrite quality and robustness.

We also report aggregated results across all datasets in Table~\ref{tab:data_efficiency}. It shows that SPA-32B consistently achieves the highest equivalence rate, the highest nontrivial speedup rate, and the highest fractions of rewrites with speedup greater than $2\times$ and $10\times$, while obtaining the lowest fractions of rewrites with slowdown greater than $4\times$ and $10\times$, as well as the lowest mean, P50, P75, P90, and P95 runtimes.

% As shown in Table~\ref{tab:main_result}, the on-policy self-improvement greatly improves performance. For Qwen3-8B trained with SPA framework, inequiv error is reduced by xx\%, xx\%, xx\%, xx\% and xx\%, nontrivial speedup rate, for TPC-H, TPC-DS, IMDB-JOB, DSB and StackOverflow, perspectively.
% Besides these, nearly all speedup percentiles was enhanced.

\subsection{Ablation Study}
\label{sec:ablation}

Table~\ref{tab:ablation_tpch_dsb} reports an ablation study of SPA-8B on TPC-H and DSB, where TPC-H represents an IID workload and DSB represents an OOD workload. Each ablated variant is compared against the full SPA-8B model with $\tau=50\%$. Overall, the full model provides the strongest latency-oriented performance across both workloads. On TPC-H, it reduces the mean runtime from $57.00$s to $38.28$s, while maintaining a $100.00\%$ success rate and a $90.91\%$ equivalence rate. On DSB, it reduces the mean runtime from $21.23$s to $13.20$s and lowers P95 latency from $94.43$s to $32.56$s, while maintaining a high success rate of $94.87\%$.

\noindent{\textbf{Effect of reinforcement learning.}}
The \textit{only SFT} variant performs worse than the full model, especially on the OOD workload. On TPC-H, it maintains a $100.00\%$ success rate but decreases the equivalence rate from $90.91\%$ to $86.36\%$ and increases the mean runtime from $38.28$s to $41.89$s. On DSB, the degradation is more significant: the success rate drops from $94.87\%$ to $61.54\%$, and the mean runtime increases from $13.20$s to $21.28$s. Although SFT alone can still produce some large speedups, it also introduces more slowdown risk. This suggests that supervised rewrite examples are insufficient for stable optimization, and execution-grounded RL is necessary to improve reliability and latency.

\noindent{\textbf{Effect of plan-divergence reward.}}
Removing the plan-divergence reward weakens latency improvement even when correctness remains competitive. On TPC-H, the mean runtime increases from $38.28$s to $42.90$s, and the $\geq 2\times$ speedup rate decreases from $9.09\%$ to $4.55\%$. On DSB, equivalence improves from $67.57\%$ to $83.78\%$, but mean runtime increases from $13.20$s to $21.49$s, and the $\geq 2\times$ speedup rate decreases from $7.69\%$ to $5.13\%$. This indicates that without an explicit incentive to change the physical execution plan, the model tends to generate safer but less execution-effective rewrites. Plan-divergence reward is therefore important for encouraging rewrites that lead to meaningful runtime improvements.

\noindent{\textbf{Effect of gated reward threshold.}}
The threshold $\tau$ controls how conservatively SPA advances to later reward components. A smaller threshold, $\tau=25\%$, is more aggressive, but it degrades both correctness and latency, especially on DSB, where equivalence drops to $55.56\%$ and mean runtime increases to $28.26$s. A larger threshold, $\tau=75\%$, is more conservative, but also underperforms the balanced setting: on TPC-H, mean runtime increases to $52.97$s, and on DSB, success drops to $84.62\%$. These results suggest that $\tau=50\%$ provides a better balance between exploration and stable optimization.

\noindent{\textbf{Effect of self-improvement.}}
Removing self-improvement hurts robustness, particularly on the OOD workload. On DSB, the mean runtime increases from $13.20$s to $21.55$s, and the $\geq 2\times$ slowdown rate increases from $2.56\%$ to $5.13\%$. The variant without self-improvement but with the same number of training steps remains worse than the full model, indicating that the benefit is not merely due to additional training. Instead, self-improvement provides useful corrective signals by training the policy to recover from its own harmful rewrites.

\noindent{\textbf{Effect of slowdown-only self-improvement.}}
Using only slowdown cases for self-improvement is also less effective than the full mixed self-improvement strategy. On TPC-H, the mean runtime increases to $52.84$s and P95 reaches $300.00$s. On DSB, the mean runtime is $20.84$s and the $\geq 2\times$ slowdown rate increases to $5.13\%$. This suggests that focusing only on negative cases can over-constrain the training signal. Mixing slowdown cases with regular training examples provides a more balanced and effective self-improvement stage.

\noindent{\textbf{Effect of plan information.}}
Removing plan information from the prompt also degrades performance. On DSB, the success rate slightly increases to $97.44\%$, but equivalence decreases to $65.79\%$, mean runtime increases to $20.96$s, and the $\geq 2\times$ slowdown rate drops to $0.00\%$ at the cost of weaker latency improvement. On TPC-H, the effect is smaller, but the variant still does not consistently outperform the full model. These results show that plan information is useful beyond syntactic guidance: it helps the model identify execution bottlenecks and generate rewrites that are more likely to improve runtime.

\noindent{\textbf{Summary.}}
The ablation results show that SPA-8B benefits from all major design choices: execution-grounded RL, plan-divergence reward, balanced gated reward progression, plan-aware prompting, and mixed self-improvement. The full model achieves the best overall latency reduction across IID and OOD workloads while maintaining high success and equivalence rates, confirming the importance of combining correctness-aware RL with execution-plan grounding.

\input{tables/speedup}
\input{tables/case_study}
\subsection{Threshold-Based Speedup and Slowdown Analysis}

Table~\ref{tab:large_magnitude_runtime} further analyzes speedup and slowdown rates under different thresholds on TPC-H and DSB. SPA-32B achieves the highest speedup rates on both datasets, and its advantage remains stable as the threshold increases from $2\times$ to $10\times$. For slowdown, SPA-32B shows low risk across both datasets. It produces no $\geq 10\times$ slowdowns on either benchmark, while keeping $\geq 2\times$ and $\geq 4\times$ slowdown rates comparable to or lower than most baselines. These results indicate that SPA improves the chance of finding substantial speedups without introducing severe slowdown cases.

\input{tables/2k}

\subsection{Case Study}
\label{sec:case_study}

Figure~\ref{fig:rewrite_comparison} presents a representative case where SPA finds an execution-effective rewrite, while both rule-based systems and a strong LLM baseline fail to obtain the same improvement.
The original query computes the total discount from \texttt{web\_sales} for items produced by manufacturer 718 within a 90-day date range.
It contains a correlated subquery that computes, for each outer item, an item-level average discount threshold over the same date range.

Although this query admits an effective decorrelation-based rewrite, both LearnedRewrite and R-Bot fail to apply any rewrite rule and leave the query unchanged.
GPT-5.4 generates an intuitive rewrite.
It correctly recognizes that the correlated average can be decorrelated into an item-level pre-aggregation, materialized as \texttt{item\_avg\_discount}, and then joined back to the outer query.
This transformation appears reasonable and is semantically valid since it returns the correct result.
However, the resulting plan exposes an execution-critical inefficiency: the planner materializes \texttt{item\_avg\_discount} via a \texttt{HashAggregate} over all 207,521 date-filtered rows from \texttt{web\_sales}, producing 50,238 distinct item groups, before the manufacturer predicate is applied.
The manufacturer filter only enters the plan as a late \texttt{Hash Join} on top of this already-computed aggregate, discarding all but 16 of the 50,238 groups.
This leads to a slowdown of $0.85\times$.

SPA produces a structurally different plan.
By joining \texttt{filtered\_} \texttt{items} (32 manufacturer-718 items) into the aggregation CTE before grouping, the planner is able to replace the broad \texttt{HashAggregate} with a \texttt{GroupAggregate} over a \texttt{Nested Loop} that first ranges over the 32 known items and retrieves their \texttt{web\_}\texttt{sales} rows via index scan.
Only 67 rows enter the aggregation, producing 16 groups—exactly the items that matter. The final join operates on this compact intermediate result, achieving a 
$28.1\times$ speedup.

The case highlights that even a strong LLM can produce a semantically valid rewrite that appears reasonable at the SQL level, but still miss the execution-critical transformation needed for speedup.

\subsection{Data Efficiency Analysis}
\label{sec:data_efficiency}

% Preparing high-quality rewritten SQL targets for SFT is the major costs of data synthesis, while original SQL queries can be generated from benchmark toolkits with nearly no cost. We therefore evaluate whether SPA remains effective when the amount of supervised rewrite data is reduced from 5k to 2k examples, and generated with only \texttt{o3-mini} model.

Preparing high-quality rewritten SQL targets for SFT is the major cost in data synthesis, whereas original SQL queries can be generated cheaply from benchmark toolkits. We therefore evaluate a lower-cost setting that reduces the SFT rewrite data from 5k to 2k examples. The 5k setting uses both benchmark-toolkit and SQLStorm-generated queries, with rewrites generated by GPT-5.4 and o3-mini. In contrast, the 2k setting uses only benchmark-toolkit queries and generates all rewrite targets with o3-mini.

Table~\ref{tab:data_efficiency} reports the data efficiency results of SPA-32B. Even with only 2k training examples, SPA-32B remains highly competitive with strong closed-source baselines. It achieves a higher equivalence rate than both GPT-5.4 and Gemini-2.5-Pro, with a mean runtime of 20.92s that is close to GPT-5.4's 19.55s and lower than Gemini-2.5-Pro's 24.95s. SPA-32B (2k) also preserves strong high-impact speedup behavior, achieving the best nontrivial speedup rate, matching Gemini-2.5-Pro at 23.95\%, while obtaining the highest ratios of queries with at least $2\times$ and $4\times$ speedup. Compared with GPT-5.4, SPA-32B (2k) also reduces nontrivial slowdown from 21.56\% to 7.78\%, and achieves lower slowdown rates across all slowdown thresholds. These results indicate that SPA can learn effective rewrite policies even with substantially reduced SFT data.

The last row further compares the full 5k setting against the 2k setting. Increasing SFT data size from 2k to 5k improves most metrics, with particularly clear gains in mean runtime and all runtime percentiles. However, the 2k model already remains competitive with strong proprietary LLMs, suggesting that SPA is relatively data-efficient while still benefiting from additional high-quality rewrite data.

\begin{table}[t]
\centering
\caption{Training cost and time by stage.}
\label{tab:training_cost_time}
\scriptsize
\setlength{\tabcolsep}{2pt}
\resizebox{\columnwidth}{!}{%
\begin{tabular}{lcccc}
\toprule
Base Model & SFT & Initial Optimization & Self-Improvement & Total \\
\midrule
Qwen3-8B  & \$2.54 / 0.13h  & \$5.38 / 9.43h  & \$5.03 / 10.88h  & \$12.95 / 20.44h \\
\midrule
Qwen3-32B & \$9.32 / 0.29h  & \$21.34 / 11.93h & \$21.42 / 12.77h & \$52.08 / 24.99h \\
\bottomrule
\end{tabular}%
}
\end{table}

\subsection{Cost Analysis}

Table~\ref{tab:training_cost_time} reports the cost and training time of each training stage.
We train all stages with LoRA adapters.
Because training is conducted through an API-based training platform, we compute cost using the platform's token-level billing scheme, which separately accounts for prefill, sampling, and training tokens.

% The SFT stage is computed over four epochs, while the RL stages include both rollout generation and policy optimization.
For Qwen3-8B, the complete training pipeline costs \$12.95 and takes 20.44 hours.
For Qwen3-32B, the total cost is \$52.08 and the total time is 24.99 hours.
The RL stages dominate both cost and time because they require sampling candidate rewrites and updating the policy using execution-based rewards.
Nevertheless, the overall training cost remains modest: even for Qwen3-32B, the full training pipeline costs around \$52.

\section{Conclusion}
We presented SPA, a SQL-Plan-Aware reinforcement learning framework for LLM-based SQL query rewriting. SPA grounds rewrite optimization via database execution by jointly considering semantic equivalence, textual rewrite distance, physical-plan divergence, and measured runtime speedup. To make execution-grounded RL effective for SQL rewriting, SPA introduces parallel reward calculation, probability-gated adaptive reward shaping, and on-policy self-improvement from slowdown-inducing rollouts. Experiments on both IID and OOD workloads show that SPA improves end-to-end runtime performance over rule-based and strong LLM baselines, while reducing trivial rewrites and harmful slowdowns. Ablation studies further demonstrate that plan-aware reward modeling, execution feedback, and self-improvement each contribute to the final performance gains. Overall, SPA shows that explicitly aligning LLM rewrite policies with physical execution behavior is a promising direction for robust and effective SQL query optimization.

% \begin{acks}
%  This work was supported by the [...] Research Fund of [...] (Number [...]). Additional funding was provided by [...] and [...]. We also thank [...] for contributing [...].
% \end{acks}

%\clearpage

\bibliographystyle{ACM-Reference-Format}
\bibliography{reference}

\end{document}

%% file: helpers/macros.tex
\definecolor{codegreen}{rgb}{0,0.6,0}
\definecolor{codegray}{rgb}{0.5,0.5,0.5}
\definecolor{codepurple}{rgb}{0.58,0,0.82}
\definecolor{darkgreen}{rgb}{0,0.5,0}
\definecolor{darkred}{rgb}{0.7, 0.1, 0.12}

\lstdefinestyle{sqlcodestyle}{
    commentstyle=\color{codegreen},
    keywordstyle=\color{codepurple},
    numberstyle=\tiny\color{codegray},
    stringstyle=\color{magenta},
    basicstyle=\ttfamily\footnotesize,
    breakatwhitespace=false,         
    breaklines=true,                 
    captionpos=b,                    
    keepspaces=true,                 
    numbersep=5pt,                  
    showspaces=false,                
    showstringspaces=false,
    showtabs=false,                  
    tabsize=2
}

\lstdefinestyle{inlinesqlcodestyle}{
    commentstyle=\color{codegreen},
    keywordstyle=\color{codepurple},
    numberstyle=\tiny\color{codegray},
    stringstyle=\color{magenta},
    basicstyle=\ttfamily\scriptsize,
    breakatwhitespace=false,         
    breaklines=true,                 
    captionpos=b,                    
    keepspaces=true,                 
    numbersep=5pt,                  
    showspaces=false,                
    showstringspaces=false,
    showtabs=false,                  
    tabsize=2
}

\newtheorem{Example}{Example}

\newcommand{\bagintersection}{{\mathbin{\ooalign{$\cap$\cr\hidewidth\raise0.2ex\hbox{\scalebox{0.6}{$+$}}\hidewidth\cr}}}}

%% file: sections/intro.tex
\section{Introduction}
Despite decades of progress in query optimization~\cite{overview,path_selection,DBLP:journals/ftdb/DingNC24}, modern DBMS optimizers remain imperfect: two semantically equivalent SQL queries can lead to substantially different execution plans and runtimes on the same system~\cite{LeisGMBK015, LeisGMBKN25}. This gap creates significant opportunities for SQL query rewriting, which transforms an input query into a semantically equivalent query that is more likely to trigger an efficient execution plan.

However, discovering effective rewrites remains challenging. Traditional rule-based rewriting techniques apply predefined transformations such as predicate pushdown, subquery decorrelation, join reordering, and common subexpression extraction~\cite{llmr2,slabcity}. These techniques are reliable and interpretable, but their effectiveness is fundamentally limited by the coverage of the rewrite rules and templates. When an optimization opportunity requires a combination of transformations that is not explicitly encoded, rule-based systems may fail to produce a beneficial rewrite.

Recent advances in large language models (LLMs) suggest a complementary direction: using LLMs to directly generate SQL rewrites~\cite{lithe,quite,genrewrite,DBLP:journals/corr/abs-2502-12918}. Unlike rule-based systems, LLMs are not restricted to a fixed transformation space and can synthesize rewrites that combine multiple structural changes. This flexibility is particularly valuable for complex analytical queries, where effective rewrites often require coordinated changes across joins, subqueries, aggregation, and filtering predicates.

\input{tables/motivation_example}

% \begin{Example}
% Consider query $Q_1$ in Figure~\ref{fig:motivation}, a variant of TPC-DS Query 1~\cite{tpcds}. The query joins \texttt{sales}, \texttt{item}, and \texttt{date}, and uses a correlated subquery to find sales whose discount is at least $30\%$ higher than the average discount for the same item over a target date range. On PostgreSQL 16 running on an EC2 m5d.2xlarge instance with 8 CPU cores and 32GB RAM, $Q_1$ takes XXX seconds to finish.

% Existing rule-based rewriting systems, including LearnedRewrite~\cite{learnedrewrite} and R-Bot~\cite{rbot}, fail to apply any beneficial Calcite~\cite{calcite} rewrite rule to this query. The missed opportunity is substantial: a modern LLM can generate the rewritten query $Q_2$, which achieves a $32.7\times$ speedup by isolating selective predicates, precomputing reusable subqueries, and, most importantly, decorrelating the expensive correlated subquery.
% \end{Example}

\begin{Example}
Consider query $Q_1$ in Figure~\ref{fig:motivation}, a variant of a TPC-DS workload~\cite{tpcds}. The query computes aggregate shipping and profit statistics for selected web-sales orders, while requiring each counted order to satisfy two additional order-level conditions: it must be associated with multiple warehouses and it must also appear in web returns. In $Q_1$, these conditions are expressed through a self-join-derived CTE that is reused inside nested \texttt{IN} subqueries. On PostgreSQL 16 running on an EC2 m5d.2xlarge instance with 8 CPU cores and 32GB RAM, $Q_1$ takes over $15$ minutes to finish.

Existing rule-based rewriting systems, including LearnedRewrite~\cite{learnedrewrite} and R-Bot~\cite{rbot}, fail to apply any Calcite~\cite{calcite} rewrite rule to this query. The missed opportunity is substantial: a modern LLM can generate the rewritten query $Q_2$, which takes only $0.53$ seconds by converting the self-join-based multi-warehouse condition into an explicit order-level aggregation, deduplicating returned orders, and reconnecting both predicates to the outer query through existence checks.
\end{Example}

This example illustrates both the promise and the risk of LLM-based SQL rewriting. On the one hand, LLMs can produce rewrites that go beyond predefined rule templates and expose optimization opportunities missed by existing systems. On the other hand, a rewrite that appears reasonable at the SQL level may still preserve the same physical plan, yield little benefit, or even lead the optimizer toward a slower execution strategy. 

This motivates treating database execution feedback as the learning signal. Rather than relying only on static rewrite examples or prompting, the model should learn from the execution outcomes of its own rewrites. SQL rewriting naturally supports this feedback loop: execution results can check semantic equivalence, query plans can reveal whether the optimizer behavior changes, and runtime measurements can quantify performance impact.

We therefore formulate SQL rewriting as a reinforcement learning problem, where the model explores candidate rewrites and receives database-grounded rewards that encourage valid, equivalent, plan-changing, and execution-effective rewrites while penalizing invalid, trivial, and harmful ones.

% This limitation points to a missing learning signal. Existing LLM-based rewriting methods largely rely on static training data or prompting: the model observes examples of SQL and rewrites, but it does not improve by observing what happens when its own rewrites are executed. Consequently, the model has no direct mechanism to learn from its mistakes: an invalid rewrite, a rewrite that preserves the original plan, and a rewrite that causes a severe slowdown are all failures whose causes are only revealed through interaction with the database system.

% SQL rewriting is particularly well suited for closing this feedback loop. Unlike many natural-language generation tasks, the quality of a rewrite can be evaluated automatically along multiple dimensions. Query execution can test semantic equivalence, query plans can reveal whether the optimizer behavior has changed, and runtime measurements can quantify performance impact. 
% These signals make SQL rewriting a particularly natural setting for reinforcement learning: the model can explore candidate rewrites, receive automatic feedback from the database system, and update its policy based on semantic and performance outcomes.

In this paper, we present \textbf{SPA}, a SQL-Plan-Aware reinforcement learning framework for LLM-based SQL query rewriting. SPA formulates SQL rewriting as an execution-grounded policy optimization problem and extends Group Relative Policy Optimization (GRPO)~\cite{grpo} with database-aware reward signals. For each generated rewrite, SPA computes rewards from four complementary aspects: semantic equivalence, textual rewrite distance, physical-plan divergence, and runtime speedup.

A key challenge is that SQL rewriting difficulty cannot be reliably inferred from static query structure alone: simple queries may have few valid rewrites, while complex ones may offer many profitable transformations. SPA addresses this with \textit{Probability-Gated Adaptive Reward Shaping (PGARS)}, a query-level curriculum that adapts the active reward target based on the policy's observed success within each rollout group. 
% PGARS organizes rewriting into hierarchical milestones (syntax validity, semantic equivalence, plan divergence, and runtime speedup), and activates higher-level rewards only after sufficient progress on preceding ones. 
This prevents the model from chasing execution gains through invalid or inequivalent rewrites while supporting stable learning across diverse query difficulties.

SPA further introduces an on-policy self-improvement procedure for learning from execution failures. Instead of treating slowdown-inducing rewrites merely as negative samples, SPA revisits them as targeted rewrite tasks for the current policy. This allows the model to learn not only which rewrites are invalid or harmful, but also how to repair them.

We evaluate SPA on both IID and OOD SQL rewriting benchmarks covering diverse analytical workloads. The results show that SPA achieves higher end-to-end query speedups than rule-based optimizers and strong LLM-based rewriting baselines. SPA also reduces both trivial rewrites and slowdown rewrites, demonstrating the benefit of plan-aware, execution-grounded policy optimization.

% Beyond aggregate performance, we provide detailed analyses of where the gains come from. Through ablation studies, threshold-based speedup and slowdown analysis, and physical-plan case studies, we show that SPA's improvements are driven by execution-grounded RL, plan-aware reward modeling, and on-policy self-improvement rather than by additional training alone.

In summary, we make the following contributions:
\begin{itemize}
\item We formulate LLM-based SQL query rewriting as an execution-grounded reinforcement learning problem, where rewrite policies are optimized using feedback from semantic validation, physical plans, and execution latency.
\item We propose \textbf{SPA}, a SQL-Plan-Aware reinforcement learning framework that trains LLMs to produce semantically valid and performance-improving SQL rewrites.
\item We introduce Probability-Gated Adaptive Reward Shaping, a query-level curriculum that progressively activates different optimization reward targets.
\item We develop an on-policy self-improvement strategy that revisits slowdown-inducing rewrites and converts execution failures into targeted training signals.
% \item We demonstrate across IID and OOD benchmarks that SPA improves runtime performance, reduces harmful slowdowns, and achieves strong tail-latency gains over existing baselines.
\item We demonstrate across IID and OOD benchmarks that SPA improves runtime performance, reduces harmful slowdowns, and achieves strong tail-latency gains over existing baselines.
\end{itemize}

%% file: tables/motivation_example.tex
\lstdefinelanguage{SQL}{
  keywords={SELECT, FROM, WHERE, JOIN, ON, GROUP, BY, WITH, AS, AND, OR, UNION, ALL, SUM, AVG, DATE, INTERVAL},
  keywordstyle=\color{green!45!black}\bfseries,
  sensitive=false,
  comment=[l]{--},
  morestring=[b]',
  stringstyle=\color{red!70!black}
}

\lstset{
  language=SQL,
  basicstyle=\ttfamily\scriptsize,
  columns=fullflexible,
  keepspaces=true,
  breaklines=true,
  showstringspaces=false,
  frame=none,
  aboveskip=0pt,
  belowskip=0pt
}

\tcbset{
  sqlbox/.style={
    enhanced,
    colback=gray!8,
    colframe=gray!55,
    boxrule=0.35pt,
    arc=1pt,
    left=4pt,
    right=4pt,
    top=3pt,
    bottom=3pt,
    before skip=3pt,
    after skip=3pt
  }
}

\begin{figure}[t]
\centering

\textbf{Original Query $Q_1$}
\begin{tcolorbox}[sqlbox]
\begin{lstlisting}
WITH multi_warehouse AS (
  SELECT ws1.order_no
  FROM web_sales ws1, web_sales ws2
  WHERE ws1.order_no = ws2.order_no
    AND ws1.wh <> ws2.wh
)
SELECT ...
FROM web_sales ws, ...
WHERE
  ... -- skipped common filters
  AND ws.order_no IN (
    SELECT order_no
    FROM multi_warehouse
  )
  AND ws.order_no IN (
    SELECT r.order_no
    FROM web_returns r, multi_warehouse m
    WHERE r.order_no = m.order_no
  );
\end{lstlisting}
\end{tcolorbox}

\vspace{2pt}

\textbf{Rewritten Query $Q_2$ (generated by GPT-5.4)}
\begin{tcolorbox}[sqlbox]
\begin{lstlisting}
WITH multi_warehouse AS (
  SELECT order_no
  FROM web_sales
  GROUP BY order_no
  HAVING COUNT(DISTINCT wh) > 1
),
returned AS (
  SELECT DISTINCT order_no
  FROM web_returns
)
SELECT ...
FROM web_sales ws
JOIN ...
WHERE
  ... --  skipped the same common filters as in Q1
  AND EXISTS (
    SELECT 1 FROM multi_warehouse m
    WHERE m.order_no = ws.order_no
  )
  AND EXISTS (
    SELECT 1 FROM returned r
    WHERE r.order_no = ws.order_no
  );
\end{lstlisting}
\end{tcolorbox}

\caption{A motivating SQL rewrite example. Common projections, joins, and filters are omitted for clarity.}
\label{fig:motivation}
\end{figure}

% to highlight the key divergence: $Q_1$ encodes the multi-warehouse condition through a self-join CTE reused in nested subqueries, while $Q_2$ turns it into an explicit order-level aggregate predicate.

%% file: related_work.tex
\section{Related Work}

\subsection{Rule-Based SQL Rewriting}
SQL query rewriting has long been supported by rule-based optimizers, which apply predefined equivalence-preserving transformations such as predicate pushdown, projection pruning, join simplification, and subquery flattening~\cite{DBLP:phd/Graefe87,DBLP:conf/vldb/LevyMS94,DBLP:conf/vldb/Muralikrishna92}. Beyond manually designed rules, WeTune~\cite{wetune} automatically discovers rewrite rules by enumerating logical query plans and verifying equivalent plan pairs with an SMT-based verifier. However, such methods primarily verify semantic correctness rather than execution effectiveness: a valid rule may still produce no speedup or even slow down execution depending on the target DBMS, statistics, data distribution, indexes, and physical plan selection. Recent learning-based approaches further improve rule application by selecting better rewrite sequences. For example, LearnedRewrite~\cite{learnedrewrite} formulates rule selection as a search problem and combines Monte Carlo Tree Search with a learned cost estimator to identify promising sequences over Calcite rules. LLM-enhanced systems further use LLMs to guide rule-based rewriting: LLM-R2~\cite{llmr2} prompts LLMs to recommend rewrite rules with contrastive demonstrations, while R-Bot~\cite{rbot} uses evidence retrieval and self-reflection to select and order rewrite rules. Although these methods improve rule discovery, selection, and ordering, their search space remains largely constrained by predefined rules, retrieved evidence, or previously observed rewrite patterns.
% In contrast, our work trains an LLM rewrite policy with database-grounded reinforcement learning feedback, enabling it to generate candidate rewrites directly and optimize for actual execution performance rather than only semantic equivalence or estimated cost.

\subsection{LLM-Based SQL Rewriting}
Recent studies have explored large language models (LLMs) for SQL query rewriting by leveraging their ability to reason over SQL semantics~\cite{DBLP:journals/pvldb/GaoWLSQDZ24,SQL-Factory,OpenSearch-SQL,DBLP:journals/pacmmod/ChenCKY25,DBLP:journals/pacmmod/YangWXWDPCL25}. Some methods use LLMs in a rule-guided manner. For example, LITHE~\cite{lithe} prompts an LLM with one selected natural-language rule from six handcrafted rewrite rules, each paired with an example, to guide the rewrite for a given query. GenRewrite~\cite{genrewrite} further uses LLMs to generate natural-language rewrite rules and apply them to SQL optimization, allowing rewrite knowledge to be expressed beyond traditional implementation-level rules. More recent agentic systems, such as QUITE~\cite{quite}, formulate rewriting as an iterative refinement process in which an LLM-based agent improves queries through multi-turn reasoning and tool interaction. These methods show that LLMs can expand SQL rewriting beyond rigid optimizer rules, but they are still largely guided by handcrafted rules, generated rule descriptions, or curated refinement actions. In contrast, our work directly trains an LLM rewrite policy with database-grounded reinforcement learning feedback, enabling the model to optimize candidate rewrites according to actual execution behavior rather than relying only on rule guidance or LLM reasoning.

\subsection{Reinforcement Learning for Databases}

Reinforcement learning (RL) has been increasingly explored for database optimization across query-level~\cite{DBLP:journals/tods/TrummerWWMMJAR21,neo,LSTM,Balsa} and system-level tasks~\cite{dba,udo}. SEFRQO~\cite{sefrqo} studies self-evolving LLM-based query optimization to continuously refine query-hint generation. For SQL generation, Reward-SQL~\cite{reward-sql} applies GRPO to align LLMs with execution-based rewards, encouraging generated SQL to be syntactically valid and executable against the target database. Beyond query-level optimization, RL has been widely used for system-level database tuning. Frameworks such as QTune~\cite{qtune} and CDBTune~\cite{cloudtune} use RL agents to automate configuration tuning and index-related decisions, demonstrating the ability of learned policies to handle complex, non-linear performance trade-offs in database systems.
E3-Rewrite~\cite{e3rewrite} proposes a GRPO framework for SQL rewriting, which builds prompts from execution plans and retrieved demonstrations. Unlike E3-Rewrite, our method does not require a demonstration pool at inference time, uses real execution feedback rather than optimizer cost estimates, and relies on database-grounded verification instead of formal-method and LLM-based equivalence judgments during training. 

%% file: tables/main_res.tex
\begin{table*}[!t]
\centering
\setlength{\tabcolsep}{2pt}
\renewcommand{\arraystretch}{1.05}
\definecolor{topone}{RGB}{161,217,155}
\definecolor{toptwo}{RGB}{199,233,192}
% Runtime highlighting helpers: best (bold + dark green) and runner-up (light green).
% Validity columns (Succ./Equiv.) are intentionally left unhighlighted.
\newcommand{\rkone}[1]{\cellcolor{topone}\textbf{#1}}
\newcommand{\rktwo}[1]{\cellcolor{toptwo}#1}

\caption{Performance comparison of SQL rewrite methods on IID and OOD workloads. TPC-H and TPC-DS are IID datasets, while DSB and StackOverflow are OOD datasets. \textbf{Succ.} denotes the fraction of generated rewrites that pass syntax checking; \textbf{Equiv.} denotes the fraction of successful rewrites that are semantically equivalent to the original query; \textbf{Mean} and \textbf{P50}/\textbf{P75}/\textbf{P90}/\textbf{P95} report the mean and percentile query runtimes. For unsuccessful or inequivalent rewrites, we apply padding by setting their runtime to the original runtime. All runtimes are in seconds; $\ge$300.00 denotes queries censored at the 300\,s per-query execution timeout. Arrows ($\uparrow$/$\downarrow$) indicate whether higher or lower is better; for each runtime metric, the \textbf{best} (bold, dark green) and second-best (light green) result is highlighted.}
\label{tab:main_results}

\resizebox{0.96\textwidth}{!}{
% \small {
\begin{tabular}{lrrrrrrrrrrrrrr}
\toprule
\makecell[l]{\textbf{IID datasets}}
& \multicolumn{7}{c}{\textbf{TPC-H (10G)}}
& \multicolumn{7}{c}{\textbf{TPC-DS (10G)}} \\
\cmidrule(lr){2-8} \cmidrule(lr){9-15}
& \multicolumn{2}{c}{\textbf{Validity}} & \multicolumn{5}{c}{\textbf{Runtime (s)}}
& \multicolumn{2}{c}{\textbf{Validity}} & \multicolumn{5}{c}{\textbf{Runtime (s)}} \\
\cmidrule(lr){2-3} \cmidrule(lr){4-8} \cmidrule(lr){9-10} \cmidrule(lr){11-15}
\textbf{Method}
& \textbf{Succ.}$\uparrow$ & \textbf{Equiv.}$\uparrow$ & \textbf{Mean}$\downarrow$ & \textbf{P50}$\downarrow$ & \textbf{P75}$\downarrow$ & \textbf{P90}$\downarrow$ & \textbf{P95}$\downarrow$
& \textbf{Succ.}$\uparrow$ & \textbf{Equiv.}$\uparrow$ & \textbf{Mean}$\downarrow$ & \textbf{P50}$\downarrow$ & \textbf{P75}$\downarrow$ & \textbf{P90}$\downarrow$ & \textbf{P95}$\downarrow$ \\
\midrule
Original
& -- & -- & 57.00 & 0.64 & 80.83 & 282.29 & $\ge$300.00
& -- & -- & 30.92 & 3.26 & 12.73 & 47.71 & $\ge$300.00 \\
% \addlinespace
\hline

LearnedRewrite
& 100.00\% & 63.64\% & 47.65 & \rktwo{0.63} & \rktwo{2.60} & 282.29 & $\ge$300.00
& 100.00\% & 35.56\% & 32.55 & 4.33 & 20.48 & 87.23 & $\ge$300.00 \\

R-Bot (Gemini-2.5-Pro)
& 95.45\% & 90.48\% & 33.19 & \rkone{0.61} & 2.64 & 95.61 & 290.27
& 82.22\% & 67.57\% & 33.06 & 3.90 & 15.84 & 91.81 & $\ge$300.00 \\
\hline
Qwen3-8B
& 100.00\% & 86.36\% & 70.00 & \rktwo{0.63} & 106.43 & $\ge$300.00 & $\ge$300.00
& 86.67\% & 78.21\% & 29.97 & 3.26 & 10.77 & 47.25 & $\ge$300.00 \\

Qwen3-32B
& 90.91\% & 90.00\% & 70.34 & 0.70 & 108.05 & $\ge$300.00 & $\ge$300.00
& 78.89\% & 74.65\% & 34.70 & 3.28 & 11.04 & 74.64 & $\ge$300.00 \\

% Gemma-4-31B-IT
% & 100.00\% & 86.36\% & 38.32 & 0.71 & 3.93 & 120.56 & 291.15
% & 100.00\% & 62.22\% & 22.89 & 2.79 & 9.68 & \rktwo{25.28} & 205.32 \\
\hline
GPT-4o
& 100.00\% & 77.27\% & 70.18 & 0.65 & 106.76 & $\ge$300.00 & $\ge$300.00
& 92.22\% & 59.04\% & 25.19 & 2.91 & 10.53 & 43.46 & 225.40 \\

Gemini-2.5-Pro
& 100.00\% & 90.91\% & \rktwo{32.94} & 0.64 & 3.77 & \rktwo{89.51} & \rktwo{289.94}
& 100.00\% & 48.89\% & 28.88 & \rktwo{2.64} & 9.80 & 32.13 & $\ge$300.00 \\

GPT-5.4
& 100.00\% & 86.36\% & 38.36 & 0.75 & 3.79 & 121.35 & 291.19
& 100.00\% & 55.56\% & \rkone{16.88} & 2.91 & \rktwo{8.50} & \rkone{24.61} & \rkone{50.12} \\
\hline
% \addlinespace
\textbf{SPA-8B}
& 100.00\% & 90.91\% & 38.28 & 0.76 & 3.73 & 120.63 & 291.15
& 93.33\% & 63.10\% & 24.09 & 2.79 & 9.56 & 31.58 & 205.32 \\

\textbf{SPA-32B}
& 100.00\% & 86.36\% & \rkone{32.73} & \rkone{0.61} & \rkone{2.36} & \rkone{89.23} & \rkone{289.93}
& 97.78\% & 68.18\% & \rktwo{19.92} & \rkone{2.63} & \rkone{7.53} & 28.46 & \rktwo{72.92} \\

\midrule
\makecell[l]{\textbf{OOD datasets}}
& \multicolumn{7}{c}{\textbf{DSB (10G)}}
& \multicolumn{7}{c}{\textbf{StackOverflow}} \\
\cmidrule(lr){2-8} \cmidrule(lr){9-15}
& \multicolumn{2}{c}{\textbf{Validity}} & \multicolumn{5}{c}{\textbf{Runtime (s)}}
& \multicolumn{2}{c}{\textbf{Validity}} & \multicolumn{5}{c}{\textbf{Runtime (s)}} \\
\cmidrule(lr){2-3} \cmidrule(lr){4-8} \cmidrule(lr){9-10} \cmidrule(lr){11-15}
\textbf{Method}
& \textbf{Succ.}$\uparrow$ & \textbf{Equiv.}$\uparrow$ & \textbf{Mean}$\downarrow$ & \textbf{P50}$\downarrow$ & \textbf{P75}$\downarrow$ & \textbf{P90}$\downarrow$ & \textbf{P95}$\downarrow$
& \textbf{Succ.}$\uparrow$ & \textbf{Equiv.}$\uparrow$ & \textbf{Mean}$\downarrow$ & \textbf{P50}$\downarrow$ & \textbf{P75}$\downarrow$ & \textbf{P90}$\downarrow$ & \textbf{P95}$\downarrow$ \\
\midrule
Original
& -- & -- & 21.23 & 2.58 & 8.67 & 17.18 & 94.43
& -- & -- & 28.12 & 0.33 & 1.87 & 71.13 & 130.31 \\
% \addlinespace
\hline
LearnedRewrite
& 100.00\% & 51.28\% & 29.17 & 3.08 & 20.78 & 74.62 & 144.12
& 100.00\% & 43.75\% & 41.05 & 0.59 & 74.88 & 130.56 & 163.74 \\

R-Bot (Gemini-2.5-Pro)
& 69.23\% & 85.19\% & 24.53 & 2.58 & 12.80 & 41.88 & 96.78
& 75.00\% & 58.33\% & 34.59 & 0.64 & 20.32 & 86.50 & 149.43 \\

Qwen3-8B
& 84.62\% & 87.88\% & 21.25 & 2.57 & 8.74 & 18.14 & 94.75
& 62.50\% & 40.00\% & 28.38 & 0.33 & 4.41 & 71.13 & 130.31 \\

Qwen3-32B
& 79.49\% & 70.97\% & 21.70 & 3.17 & 9.75 & 22.46 & 94.43
& 81.25\% & 69.23\% & 42.57 & 0.30 & 1.88 & 186.88 & $\ge$300.00 \\

% Gemma-4-31B-IT
% & 100.00\% & 87.18\% & 21.78 & 2.32 & 7.02 & 31.92 & 94.76
% & 100.00\% & 93.75\% & 10.00 & \rkone{0.19} & 4.81 & 37.96 & 69.40 \\
\hline
GPT-4o
& 89.74\% & 71.43\% & 13.75 & \rkone{1.73} & 6.80 & \rktwo{16.34} & 34.60
& 93.75\% & 53.33\% & 25.60 & 0.30 & 1.89 & 51.18 & 126.38 \\

Gemini-2.5-Pro
& 100.00\% & 87.18\% & 19.55 & \rktwo{1.74} & \rkone{5.15} & 17.18 & 56.21
& 100.00\% & 68.75\% & \rkone{5.08} & 0.22 & \rkone{0.46} & \rkone{2.68} & \rkone{21.07} \\

GPT-5.4
& 100.00\% & 82.05\% & 20.98 & 2.82 & 8.38 & 17.78 & 83.35
& 100.00\% & 75.00\% & \rktwo{5.21} & 0.22 & \rkone{0.46} & 4.65 & 22.15 \\
\hline
% \addlinespace
\textbf{SPA-8B}
& 94.87\% & 67.57\% & \rktwo{13.20} & 1.84 & \rktwo{5.71} & \rkone{13.32} & \rktwo{32.56}
& 81.25\% & 100.00\% & 5.80 & \rktwo{0.21} & 1.46 & 6.61 & 24.93 \\

\textbf{SPA-32B}
& 97.44\% & 84.21\% & \rkone{6.96} & \rktwo{1.74} & 6.67 & \rkone{13.32} & \rkone{27.85}
& 100.00\% & 100.00\% & 5.41 & 0.22 & \rktwo{1.30} & \rktwo{3.71} & \rktwo{21.63} \\
\bottomrule
\end{tabular}
}
\end{table*}

%% file: tables/ablation.tex
% required packages:
% \usepackage{booktabs}
% \usepackage{makecell}
% \usepackage{xcolor}
% \usepackage{graphicx}

% Higher is better: Succ., Equiv., Speedup
\newcommand{\hibetter}[1]{#1\,{\color{green!45!black}\scriptsize$\uparrow$}}
\newcommand{\hiworse}[1]{#1\,{\color{red!65!black}\scriptsize$\downarrow$}}

% Lower is better: runtime and slowdown
\newcommand{\lobetter}[1]{#1\,{\color{green!45!black}\scriptsize$\downarrow$}}
\newcommand{\loworse}[1]{#1\,{\color{red!65!black}\scriptsize$\uparrow$}}

% Same as full SPA
\newcommand{\same}[1]{#1\,{\color{black!45}\scriptsize--}}

\begin{table*}[!t]
\centering
\setlength{\tabcolsep}{2.0pt}
\renewcommand{\arraystretch}{1.12}
% \footnotesize
\caption{Ablation study of SPA-8B on TPC-H and DSB. TPC-H is an IID workload and DSB is an OOD workload. Succ. denotes syntax-success rate, and Equiv. denotes semantic-equivalence rate among successful rewrites. Mean/P90/P95 report padded runtime in seconds. Spd. $\geq 2\times$ and Sld. $\geq 2\times$ report large speedup and slowdown rates. Arrows indicate whether each ablated variant improves or degrades relative to Full SPA-8B under the corresponding metric direction; higher is better for success, equivalence, and speedup, while lower is better for runtime and slowdown.}
\label{tab:ablation_tpch_dsb}

\resizebox{\textwidth}{!}{
% {
\begin{tabular}{lccccccc|ccccccc}
\toprule
\multirow{2}{*}{\textbf{Variant}}
& \multicolumn{7}{c|}{\textbf{TPC-H (IID)}}
& \multicolumn{7}{c}{\textbf{DSB (OOD)}} \\
\cmidrule(lr){2-8} \cmidrule(lr){9-15}
& \makecell{\textbf{Succ.}\textbf{(\%)}}
& \makecell{\textbf{Equiv.}\textbf{(\%)}}
& \makecell{\textbf{Mean}\textbf{(s)}}
& \makecell{\textbf{P90}\textbf{(s)}}
& \makecell{\textbf{P95}\textbf{(s)}}
& \makecell{\textbf{Spd.}\textbf{$\geq 2\times$}}
& \makecell{\textbf{Sld.}\textbf{$\geq 2\times$}}
& \makecell{\textbf{Succ.}\textbf{(\%)}}
& \makecell{\textbf{Equiv.}\textbf{(\%)}}
& \makecell{\textbf{Mean}\textbf{(s)}}
& \makecell{\textbf{P90}\textbf{(s)}}
& \makecell{\textbf{P95}\textbf{(s)}}
& \makecell{\textbf{Spd.}\textbf{$\geq 2\times$}}
& \makecell{\textbf{Sld.}\textbf{$\geq 2\times$}} \\
\midrule
Original
& -- & -- & 57.00 & 282.29 & 300.00 & -- & --
& -- & -- & 21.23 & 17.18 & 94.43 & -- & -- \\
\midrule
\textbf{Full SPA-8B}
& \textbf{100.00} & \textbf{90.91} & \textbf{38.28} & \textbf{120.63} & \textbf{291.15} & \textbf{9.09} & \textbf{9.09}
& \textbf{94.87} & \textbf{67.57} & \textbf{13.20} & \textbf{13.32} & \textbf{32.56} & \textbf{7.69} & \textbf{2.56} \\
\midrule
Only SFT
& \same{100.00} & \hiworse{86.36} & \loworse{41.89} & \lobetter{107.95} & \lobetter{290.45} & \same{9.09} & \same{9.09}
& \hiworse{61.54} & \hibetter{75.00} & \loworse{21.28} & \loworse{25.52} & \loworse{94.43} & \hibetter{10.26} & \loworse{5.13} \\
\midrule
w/o plan-divergence reward
& \hiworse{95.45} & \hiworse{90.48} & \loworse{42.90} & \loworse{121.82} & \loworse{291.16} & \hiworse{4.55} & \lobetter{4.55}
& \same{94.87} & \hibetter{83.78} & \loworse{21.49} & \loworse{19.49} & \loworse{94.49} & \hiworse{5.13} & \same{2.56} \\

$\tau=25\%$
& \same{100.00} & \hiworse{86.36} & \loworse{43.19} & \loworse{121.48} & \lobetter{291.14} & \hiworse{4.55} & \same{9.09}
& \hiworse{92.31} & \hiworse{55.56} & \loworse{28.26} & \loworse{36.82} & \loworse{300.00} & \hiworse{5.13} & \loworse{5.13} \\

$\tau=75\%$
& \same{100.00} & \same{90.91} & \loworse{52.97} & \loworse{282.32} & \loworse{300.00} & \hiworse{4.55} & \same{9.09}
& \hiworse{84.62} & \hiworse{63.64} & \loworse{21.18} & \loworse{17.18} & \loworse{94.43} & \hiworse{5.13} & \same{2.56} \\
\midrule
w/o self-improvement
& \same{100.00} & \hiworse{86.36} & \loworse{38.65} & \loworse{123.59} & \loworse{291.28} & \same{9.09} & \lobetter{4.55}
& \same{94.87} & \hibetter{78.38} & \loworse{21.55} & \loworse{18.37} & \loworse{95.11} & \same{7.69} & \loworse{5.13} \\

w/o self-improvement, same steps
& \hiworse{95.45} & \hibetter{95.24} & \loworse{38.49} & \loworse{121.15} & \same{291.15} & \same{9.09} & \lobetter{4.55}
& \hiworse{89.74} & \hibetter{77.14} & \loworse{21.41} & \loworse{17.18} & \loworse{94.48} & \hiworse{2.56} & \loworse{5.13} \\

slowdown-only self-improvement
& \same{100.00} & \same{90.91} & \loworse{52.84} & \loworse{282.28} & \loworse{300.00} & \hiworse{4.55} & \same{9.09}
& \hiworse{89.74} & \hibetter{68.57} & \loworse{20.84} & \loworse{15.01} & \loworse{94.70} & \same{7.69} & \loworse{5.13} \\
\midrule
w/o plan information
& \same{100.00} & \same{90.91} & \lobetter{38.17} & \loworse{122.61} & \loworse{291.26} & \same{9.09} & \lobetter{0.00}
& \hibetter{97.44} & \hiworse{65.79} & \loworse{20.96} & \loworse{19.40} & \loworse{94.43} & \same{7.69} & \lobetter{0.00} \\
\bottomrule
\end{tabular}
}
\end{table*}

%% file: tables/speedup.tex
\begin{table}[t]
\centering
\scriptsize
\setlength{\tabcolsep}{3.2pt}
\renewcommand{\arraystretch}{1.08}

\definecolor{topone}{RGB}{161,217,155}

\caption{Speedup and slowdown ratio under different thresholds. Top-1 values are highlighted in green.}
\label{tab:large_magnitude_runtime}

\resizebox{\columnwidth}{!}{
\begin{tabular}{lcccccc}
\toprule
\textbf{Method}
& \makecell{\textbf{Spd.}\\$\geq 2\times$}
& \makecell{\textbf{Spd.}\\$\geq 4\times$}
& \makecell{\textbf{Spd.}\\$\geq 10\times$}
& \makecell{\textbf{Sld.}\\$\geq 2\times$}
& \makecell{\textbf{Sld.}\\$\geq 4\times$}
& \makecell{\textbf{Sld.}\\$\geq 10\times$} \\
\midrule

\multicolumn{7}{c}{\textbf{TPC-H}} \\
\midrule
LearnedRewrite
& 9.09\%
& 9.09\%
& 9.09\%
& \cellcolor{topone}0.00\%
& \cellcolor{topone}0.00\%
& \cellcolor{topone}0.00\% \\

R-Bot (Gemini-2.5-Pro)
& \cellcolor{topone}13.64\%
& \cellcolor{topone}13.64\%
& \cellcolor{topone}13.64\%
& 9.09\%
& 4.55\%
& \cellcolor{topone}0.00\% \\

Qwen3-8B
& 0.00\%
& 0.00\%
& 0.00\%
& 4.55\%
& 4.55\%
& 4.55\% \\

Qwen3-32B
& 0.00\%
& 0.00\%
& 0.00\%
& 9.09\%
& 4.55\%
& 4.55\% \\

% Gemma-4-31B-IT
% & 9.09\%
% & 9.09\%
% & 9.09\%
% & 9.09\%
% & 4.55\%
% & \cellcolor{topone}0.00\% \\

GPT-4o
& 0.00\%
& 0.00\%
& 0.00\%
& 9.09\%
& 9.09\%
& 4.55\% \\

Gemini-2.5-Pro
& \cellcolor{topone}13.64\%
& \cellcolor{topone}13.64\%
& \cellcolor{topone}13.64\%
& 13.64\%
& 9.09\%
& 4.55\% \\

GPT-5.4
& 9.09\%
& 9.09\%
& 9.09\%
& 9.09\%
& 4.55\%
& \cellcolor{topone}0.00\% \\

SPA-8B
& 9.09\%
& 9.09\%
& 9.09\%
& 9.09\%
& 4.55\%
& \cellcolor{topone}0.00\% \\

\textbf{SPA-32B}
& \cellcolor{topone}\textbf{13.64\%}
& \cellcolor{topone}\textbf{13.64\%}
& \cellcolor{topone}\textbf{13.64\%}
& 9.09\%
& 4.55\%
& \cellcolor{topone}\textbf{0.00\%} \\

\midrule
\multicolumn{7}{c}{\textbf{DSB}} \\
\midrule
LearnedRewrite
& \cellcolor{topone}10.26\%
& 5.13\%
& 5.13\%
& 12.82\%
& 12.82\%
& 12.82\% \\

R-Bot (Gemini-2.5-Pro)
& 0.00\%
& 0.00\%
& 0.00\%
& 10.26\%
& 5.13\%
& 2.56\% \\

Qwen3-8B
& 0.00\%
& 0.00\%
& 0.00\%
& \cellcolor{topone}0.00\%
& \cellcolor{topone}0.00\%
& \cellcolor{topone}0.00\% \\

Qwen3-32B
& 0.00\%
& 0.00\%
& 0.00\%
& 5.13\%
& \cellcolor{topone}0.00\%
& \cellcolor{topone}0.00\% \\

% Gemma-4-31B-IT
% & 5.13\%
% & 5.13\%
% & 0.00\%
% & 5.13\%
% & 2.56\%
% & \cellcolor{topone}0.00\% \\

GPT-4o
& 7.69\%
& 7.69\%
& 2.56\%
& 2.56\%
& \cellcolor{topone}0.00\%
& \cellcolor{topone}0.00\% \\

Gemini-2.5-Pro
& \cellcolor{topone}10.26\%
& 5.13\%
& 2.56\%
& 10.26\%
& 5.13\%
& \cellcolor{topone}0.00\% \\

GPT-5.4
& 7.69\%
& 7.69\%
& 2.56\%
& 17.95\%
& 12.82\%
& \cellcolor{topone}0.00\% \\

SPA-8B
& 7.69\%
& 5.13\%
& 2.56\%
& 2.56\%
& \cellcolor{topone}0.00\%
& \cellcolor{topone}0.00\% \\

\textbf{SPA-32B}
& \cellcolor{topone}\textbf{10.26\%}
& \cellcolor{topone}\textbf{10.26\%}
& \cellcolor{topone}\textbf{7.69\%}
& 7.69\%
& 2.56\%
& \cellcolor{topone}\textbf{0.00\%} \\

\bottomrule
\end{tabular}
}
\end{table}

%% file: tables/case_study.tex
\lstdefinelanguage{SQL}{
  keywords={
    SELECT,FROM,WHERE,JOIN,ON,GROUP,BY,WITH,AS,AND,OR,
    SUM,AVG,BETWEEN
  },
  keywordstyle=\color{green!40!black}\bfseries,
  sensitive=false,
  comment=[l]{--},
  commentstyle=\color{gray}\itshape,
  morestring=[b]',
  stringstyle=\color{red!70!black}
}

\lstset{
  language=SQL,
  basicstyle=\ttfamily\fontsize{5.8pt}{6.4pt}\selectfont,
  columns=fullflexible,
  keepspaces=true,
  breaklines=true,
  showstringspaces=false,
  frame=none,
  aboveskip=0pt,
  belowskip=0pt,
  lineskip=-1pt
}

\tcbset{
  sqlbox/.style={
    enhanced,
    colback=gray!5,
    colframe=black!45,
    boxrule=0.35pt,
    arc=0pt,
    left=2pt,
    right=2pt,
    top=2pt,
    bottom=2pt,
    width=\linewidth,
    height=0.36\textheight,
    valign=top,
    before skip=0pt,
    after skip=0pt
  }
}

\begin{figure*}[t]
\centering
\setlength{\tabcolsep}{3pt}

\begin{tabular}{@{}p{0.31\textwidth} p{0.33\textwidth} p{0.33\textwidth}@{}}

\begin{minipage}[t]{\linewidth}
\vspace{0pt}
\begin{tcolorbox}[sqlbox]
\begin{lstlisting}[language=SQL]
-- original query
SELECT
    SUM(ws.discount)
FROM
    web_sales ws,
    item i,
    date_dim d
WHERE
    i.manufact_id = 718
    AND i.item_id = ws.item_id
    AND d.date_id = ws.sold_date_id
    AND d.date BETWEEN ...
    AND ws.discount > (
        SELECT
            1.3 * AVG(ws2.discount)
        FROM
            web_sales ws2,
            date_dim d2
        WHERE
            ws2.item_id = i.item_id
            AND d2.date_id = ws2.sold_date_id
            AND d2.date BETWEEN ...
    );
\end{lstlisting}
\end{tcolorbox}
\end{minipage}

&

\begin{minipage}[t]{\linewidth}
\vspace{0pt}
\begin{tcolorbox}[sqlbox]
\begin{lstlisting}[language=SQL]
-- rewritten query by GPT-5.4 (speedup=0.85)
WITH dates AS (
    SELECT date_id
    FROM date_dim
    WHERE date BETWEEN ...
),
items AS (
    SELECT item_id
    FROM item
    WHERE manufact_id = 718
),
item_avg_discount AS (
    SELECT
        ws.item_id,
        1.3 * AVG(ws.discount) AS threshold
    FROM
        web_sales ws
        JOIN dates d
          ON d.date_id = ws.sold_date_id
    GROUP BY
        ws.item_id
)
SELECT
    SUM(ws.discount)
FROM
    web_sales ws
    JOIN dates d
      ON d.date_id = ws.sold_date_id
    JOIN items i
      ON i.item_id = ws.item_id
    JOIN item_avg_discount a
      ON a.item_id = ws.item_id
WHERE
    ws.discount > a.threshold;
\end{lstlisting}
\end{tcolorbox}
\end{minipage}

&

\begin{minipage}[t]{\linewidth}
\vspace{0pt}
\begin{tcolorbox}[sqlbox]
\begin{lstlisting}[language=SQL]
-- rewritten query by SPA (speedup=28.1)
WITH items AS (
    SELECT item_id
    FROM item
    WHERE manufact_id = 718
),
item_avg_discount AS (
    SELECT
        ws.item_id,
        1.3 * AVG(ws.discount) AS threshold
    FROM
        web_sales ws
        JOIN date_dim d ON ...
        JOIN items i
          ON i.item_id = ws.item_id
    WHERE
        d.date BETWEEN ...
    GROUP BY
        ws.item_id
)
SELECT
    SUM(ws.discount)
FROM
    web_sales ws
    JOIN items i
      ON i.item_id = ws.item_id
    JOIN date_dim d ON ...
    JOIN item_avg_discount a
      ON a.item_id = ws.item_id
WHERE
    d.date BETWEEN ...
    AND ws.discount > a.threshold;
\end{lstlisting}
\end{tcolorbox}
\end{minipage}

\end{tabular}

\caption{Simplified comparison of the original query and two rewrites generated by GPT-5.4 and SPA. The original query is a variation of Query-10 from TPC-DS benchmark.}
\label{fig:rewrite_comparison}
\end{figure*}

%% file: tables/2k.tex
\begin{table*}[t]
\centering
\scriptsize
\setlength{\tabcolsep}{3.5pt}
\renewcommand{\arraystretch}{1.08}

\definecolor{topone}{RGB}{161,217,155}
\definecolor{toptwo}{RGB}{199,233,192}
\definecolor{gooddelta}{RGB}{35,139,69}
\definecolor{baddelta}{RGB}{203,24,29}

\caption{Data efficiency analysis of SPA-32B, with results aggregated across all datasets. Top-1 and Top-2 values are highlighted in green, excluding the 5k setting. The last row reports the 5k SFT data setting with changes relative to the 2k setting.}

\label{tab:data_efficiency}

\resizebox{\textwidth}{!}{
\begin{tabular}{lccccccccccccccc}
\toprule
\textbf{Method}
& \makecell{\textbf{Succ.}\\\textbf{Rate}}
& \makecell{\textbf{Equiv.}\\\textbf{Rate}}
& \textbf{Mean}
& \textbf{P50}
& \textbf{P75}
& \textbf{P90}
& \textbf{P95}
& \makecell{\textbf{Nontriv.}\\\textbf{Spd.}}
& \makecell{\textbf{Spd.}\\$\geq 2\times$}
& \makecell{\textbf{Spd.}\\$\geq 4\times$}
& \makecell{\textbf{Spd.}\\$\geq 10\times$}
& \makecell{\textbf{Nontriv.}\\\textbf{Sld.}}
& \makecell{\textbf{Sld.}\\$\geq 2\times$}
& \makecell{\textbf{Sld.}\\$\geq 4\times$}
& \makecell{\textbf{Sld.}\\$\geq 10\times$} \\
\midrule

Original
& -- & --
& 31.82 & 2.35 & 10.87 & 89.61 & 300.00
& -- & -- & -- & --
& -- & -- & -- & -- \\

LearnedRewrite
& \cellcolor{topone}100.0\% & 43.71\%
& 34.56 & 2.71 & 21.80 & 101.52 & 300.00
& 8.38\% & 6.59\% & 5.39\% & 5.39\%
& 20.96\% & 15.57\% & 11.98\% & 9.58\% \\

R-Bot (Gemini-2.5-Pro)
& 80.24\% & 73.88\%
& 31.23 & 2.75 & 13.41 & 80.36 & 300.00
& 6.59\% & 2.99\% & 2.99\% & 2.40\%
& 16.17\% & 10.18\% & 5.99\% & 4.19\% \\

Qwen3-32B
& 80.84\% & \cellcolor{toptwo}75.56\%
& 37.11 & 2.61 & 10.99 & 114.49 & 300.00
& 9.58\% & 2.40\% & 0.00\% & 0.00\%
& 10.18\% & 5.99\% & \cellcolor{toptwo}2.99\% & 2.40\% \\

GPT-4o
& 92.81\% & 63.87\%
& 28.49 & \cellcolor{topone}1.74 & 9.97 & 69.90 & 300.00
& 10.18\% & 6.59\% & 3.59\% & 2.40\%
& \cellcolor{topone}7.19\% & \cellcolor{topone}2.99\% & \cellcolor{topone}1.80\% & 1.20\% \\

% Gemma-31B-IT
% & \cellcolor{topone}100.0\% & 74.25\%
% & 23.43 & 1.99 & 7.55 & 39.88 & 246.90
% & 20.36\% & 11.38\% & 9.58\% & 5.99\%
% & 8.98\% & \cellcolor{toptwo}3.59\% & \cellcolor{topone}1.80\% & \cellcolor{topone}0.00\% \\

Gemini-2.5-Pro
& \cellcolor{topone}100.0\% & 65.27\%
& 24.95 & \cellcolor{toptwo}1.74 & \cellcolor{topone}5.74 & \cellcolor{toptwo}28.52 & 300.00
& \cellcolor{topone}23.95\% & \cellcolor{toptwo}13.17\% & \cellcolor{toptwo}10.18\% & \cellcolor{toptwo}6.59\%
& 12.57\% & 6.59\% & 3.59\% & 1.20\% \\

GPT-5.4
& \cellcolor{topone}100.0\% & 67.66\%
& \cellcolor{topone}19.55 & 1.98 & 7.36 & \cellcolor{topone}25.31 & \cellcolor{topone}95.80
& \cellcolor{toptwo}21.56\% & \cellcolor{toptwo}13.17\% & \cellcolor{toptwo}10.18\% & \cellcolor{topone}8.38\%
& 21.56\% & 13.77\% & 6.59\% & 1.80\% \\

\textbf{SPA-32B (2k)}
& \cellcolor{toptwo}\textbf{97.60\%} & \cellcolor{topone}\textbf{77.30\%}
& \cellcolor{toptwo}\textbf{20.92} & \textbf{1.79} & \cellcolor{toptwo}\textbf{6.13} & \textbf{30.83} & \cellcolor{toptwo}\textbf{112.78}
& \cellcolor{topone}\textbf{23.95\%} & \cellcolor{topone}\textbf{14.37\%} & \cellcolor{topone}\textbf{11.38\%} & \cellcolor{toptwo}\textbf{6.59\%}
& \cellcolor{toptwo}\textbf{7.78\%} & \cellcolor{topone}\textbf{2.99\%} & \cellcolor{topone}\textbf{1.80\%} & \cellcolor{toptwo}\textbf{0.60\%} \\

\midrule

\textbf{SPA-32B (5k)}
& \makecell{\textbf{98.20\%}\\{\scriptsize\textcolor{gooddelta}{+0.60\%}}}
& \makecell{\textbf{77.44\%}\\{\scriptsize\textcolor{gooddelta}{+0.14\%}}}
& \makecell{\textbf{17.19}\\{\scriptsize\textcolor{gooddelta}{-3.73}}}
& \makecell{\textbf{1.69}\\{\scriptsize\textcolor{gooddelta}{-0.10}}}
& \makecell{\textbf{5.93}\\{\scriptsize\textcolor{gooddelta}{-0.21}}}
& \makecell{\textbf{26.14}\\{\scriptsize\textcolor{gooddelta}{-4.70}}}
& \makecell{\textbf{73.10}\\{\scriptsize\textcolor{gooddelta}{-39.68}}}
& \makecell{\textbf{23.95\%}\\{\scriptsize\textcolor{gooddelta}{+0.00\%}}}
& \makecell{\textbf{14.97\%}\\{\scriptsize\textcolor{gooddelta}{+0.60\%}}}
& \makecell{\textbf{10.78\%}\\{\scriptsize\textcolor{baddelta}{-0.60\%}}}
& \makecell{\textbf{8.98\%}\\{\scriptsize\textcolor{gooddelta}{+2.39\%}}}
& \makecell{\textbf{13.17\%}\\{\scriptsize\textcolor{baddelta}{+5.39\%}}}
& \makecell{\textbf{5.99\%}\\{\scriptsize\textcolor{baddelta}{+3.00\%}}}
& \makecell{\textbf{1.80\%}\\{\scriptsize\textcolor{gooddelta}{+0.00\%}}}
& \makecell{\textbf{0.00\%}\\{\scriptsize\textcolor{gooddelta}{-0.60\%}}} \\

\bottomrule
\end{tabular}
}
\end{table*}

%% file: reference.bib
@article{learnedrewrite,
  author       = {Xuanhe Zhou and
                  Guoliang Li and
                  Chengliang Chai and
                  Jianhua Feng},
  title        = {A Learned Query Rewrite System using Monte Carlo Tree Search},
  journal      = {Proc. {VLDB} Endow.},
  volume       = {15},
  number       = {1},
  pages        = {46--58},
  year         = {2021},
  url          = {http://www.vldb.org/pvldb/vol15/p46-li.pdf},
  doi          = {10.14778/3485450.3485456},
  bibsource    = {dblp computer science bibliography, https://dblp.org}
}

@inproceedings{lithe,
  author       = {Sriram Dharwada and
                  Himanshu Devrani and
                  Jayant R. Haritsa and
                  Harish Doraiswamy},
  editor       = {Wolfgang Lehner and
                  Vanessa Braganholo and
                  Kostas Stefanidis and
                  Zheying Zhang and
                  Alexander Krause and
                  Jo{\~{a}}o Felipe Nicolaci Pimentel},
  title        = {{LITHE:} {A} Query Rewrite Advisor using LLMs},
  booktitle    = {Proceedings 29th International Conference on Extending Database Technology,
                  {EDBT} 2026, Tampere, Finland, March 24-27, 2026},
  pages        = {233--246},
  publisher    = {OpenProceedings.org},
  year         = {2026},
  url          = {https://doi.org/10.48786/edbt.2026.20},
  doi          = {10.48786/EDBT.2026.20},
  bibsource    = {dblp computer science bibliography, https://dblp.org}
}

@inproceedings{wetune,
  author       = {Zhaoguo Wang and
                  Zhou Zhou and
                  Yicun Yang and
                  Haoran Ding and
                  Gansen Hu and
                  Ding Ding and
                  Chuzhe Tang and
                  Haibo Chen and
                  Jinyang Li},
  editor       = {Zachary G. Ives and
                  Angela Bonifati and
                  Amr El Abbadi},
  title        = {WeTune: Automatic Discovery and Verification of Query Rewrite Rules},
  booktitle    = {{SIGMOD} '22: International Conference on Management of Data, Philadelphia,
                  PA, USA, June 12 - 17, 2022},
  pages        = {94--107},
  publisher    = {{ACM}},
  year         = {2022},
  url          = {https://doi.org/10.1145/3514221.3526125},
  doi          = {10.1145/3514221.3526125},
  bibsource    = {dblp computer science bibliography, https://dblp.org}
}

@article{llmr2,
author = {Li, Zhaodonghui and Yuan, Haitao and Wang, Huiming and Cong, Gao and Bing, Lidong},
title = {LLM-R2: A Large Language Model Enhanced Rule-Based Rewrite System for Boosting Query Efficiency},
year = {2024},
issue_date = {September 2024},
publisher = {VLDB Endowment},
volume = {18},
number = {1},
issn = {2150-8097},
url = {https://doi.org/10.14778/3696435.3696440},
doi = {10.14778/3696435.3696440},
journal = {Proc. VLDB Endow.},
month = sep,
pages = {53–65},
numpages = {13}
}

@article{quite,
  author       = {Yuyang Song and
                  Hanxu Yan and
                  Jiale Lao and
                  Yibo Wang and
                  Yufei Li and
                  Yuanchun Zhou and
                  Jianguo Wang and
                  Mingjie Tang},
  title        = {{QUITE:} {A} Query Rewrite System Beyond Rules with {LLM} Agents},
  journal      = {CoRR},
  volume       = {abs/2506.07675},
  year         = {2025},
  url          = {https://doi.org/10.48550/arXiv.2506.07675},
  doi          = {10.48550/ARXIV.2506.07675},
  eprinttype   = {arXiv},
  eprint       = {2506.07675},
  bibsource    = {dblp computer science bibliography, https://dblp.org}
}

@article{genrewrite,
  title={GenRewrite: Query Rewriting via Large Language Models},
  author={Liu, Jie and Mozafari, Barzan},
  journal={Proceedings of the ACM on Management of Data},
  volume={4},
  number={1 (SIGMOD},
  pages={1--26},
  year={2026},
  publisher={ACM New York, NY, USA}
}

@article{reward-sql,
  author       = {Yuxin Zhang and
                  Meihao Fan and
                  Ju Fan and
                  Mingyang Yi and
                  Yuyu Luo and
                  Jian Tan and
                  Guoliang Li},
  title        = {Reward-SQL: Boosting Text-to-SQL via Stepwise Reasoning and Process-Supervised
                  Rewards},
  journal      = {CoRR},
  volume       = {abs/2505.04671},
  year         = {2025},
  url          = {https://doi.org/10.48550/arXiv.2505.04671},
  doi          = {10.48550/ARXIV.2505.04671},
  eprinttype   = {arXiv},
  eprint       = {2505.04671},
  bibsource    = {dblp computer science bibliography, https://dblp.org}
}

@article{qtune,
author = {Li, Guoliang and Zhou, Xuanhe and Li, Shifu and Gao, Bo},
title = {QTune: a query-aware database tuning system with deep reinforcement learning},
year = {2019},
issue_date = {August 2019},
publisher = {VLDB Endowment},
volume = {12},
number = {12},
issn = {2150-8097},
url = {https://doi.org/10.14778/3352063.3352129},
doi = {10.14778/3352063.3352129},
journal = {Proc. VLDB Endow.},
month = aug,
pages = {2118–2130},
numpages = {13}
}

@inproceedings{cloudtune,
author = {Zhang, Ji and Liu, Yu and Zhou, Ke and Li, Guoliang and Xiao, Zhili and Cheng, Bin and Xing, Jiashu and Wang, Yangtao and Cheng, Tianheng and Liu, Li and Ran, Minwei and Li, Zekang},
title = {An End-to-End Automatic Cloud Database Tuning System Using Deep Reinforcement Learning},
year = {2019},
isbn = {9781450356435},
publisher = {Association for Computing Machinery},
address = {New York, NY, USA},
url = {https://doi.org/10.1145/3299869.3300085},
doi = {10.1145/3299869.3300085},
booktitle = {Proceedings of the 2019 International Conference on Management of Data},
pages = {415–432},
numpages = {18},
location = {Amsterdam, Netherlands},
series = {SIGMOD '19}
}

@article{SQLStorm,
  author       = {Tobias Schmidt and
                  Viktor Leis and
                  Peter Boncz and
                  Thomas Neumann},
  title        = {SQLStorm: Taking Database Benchmarking into the {LLM} Era},
  journal      = {Proc. {VLDB} Endow.},
  volume       = {18},
  number       = {11},
  pages        = {4144--4157},
  year         = {2025},
  url          = {https://www.vldb.org/pvldb/vol18/p4144-schmidt.pdf},
  doi          = {10.14778/3749646.3749683},
  bibsource    = {dblp computer science bibliography, https://dblp.org}
}

@article{dsb,
  author       = {Bailu Ding and
                  Surajit Chaudhuri and
                  Johannes Gehrke and
                  Vivek R. Narasayya},
  title        = {{DSB:} {A} Decision Support Benchmark for Workload-Driven and Traditional
                  Database Systems},
  journal      = {Proc. {VLDB} Endow.},
  volume       = {14},
  number       = {13},
  pages        = {3376--3388},
  year         = {2021},
  url          = {http://www.vldb.org/pvldb/vol14/p3376-ding.pdf},
  doi          = {10.14778/3484224.3484234},
  bibsource    = {dblp computer science bibliography, https://dblp.org}
}

@inproceedings{tpcds,
  author       = {Raghunath Othayoth Nambiar and
                  Meikel Poess},
  editor       = {Umeshwar Dayal and
                  Kyu{-}Young Whang and
                  David B. Lomet and
                  Gustavo Alonso and
                  Guy M. Lohman and
                  Martin L. Kersten and
                  Sang Kyun Cha and
                  Young{-}Kuk Kim},
  title        = {The Making of {TPC-DS}},
  booktitle    = {Proceedings of the 32nd International Conference on Very Large Data
                  Bases, Seoul, Korea, September 12-15, 2006},
  pages        = {1049--1058},
  publisher    = {{ACM}},
  year         = {2006},
  url          = {http://dl.acm.org/citation.cfm?id=1164217},
  bibsource    = {dblp computer science bibliography, https://dblp.org}
}

@inproceedings{bao,
  author       = {Ryan Marcus and
                  Parimarjan Negi and
                  Hongzi Mao and
                  Nesime Tatbul and
                  Mohammad Alizadeh and
                  Tim Kraska},
  editor       = {Guoliang Li and
                  Zhanhuai Li and
                  Stratos Idreos and
                  Divesh Srivastava},
  title        = {Bao: Making Learned Query Optimization Practical},
  booktitle    = {{SIGMOD} '21: International Conference on Management of Data, Virtual
                  Event, China, June 20-25, 2021},
  pages        = {1275--1288},
  publisher    = {{ACM}},
  year         = {2021},
  url          = {https://doi.org/10.1145/3448016.3452838},
  doi          = {10.1145/3448016.3452838},
  bibsource    = {dblp computer science bibliography, https://dblp.org}
}

@inproceedings{lora,
  author       = {Edward J. Hu and
                  Yelong Shen and
                  Phillip Wallis and
                  Zeyuan Allen{-}Zhu and
                  Yuanzhi Li and
                  Shean Wang and
                  Lu Wang and
                  Weizhu Chen},
  title        = {LoRA: Low-Rank Adaptation of Large Language Models},
  booktitle    = {The Tenth International Conference on Learning Representations, {ICLR}
                  2022, Virtual Event, April 25-29, 2022},
  publisher    = {OpenReview.net},
  year         = {2022},
  url          = {https://openreview.net/forum?id=nZeVKeeFYf9},
  bibsource    = {dblp computer science bibliography, https://dblp.org}
}

@article{tpch,
author = {Poess, Meikel and Floyd, Chris},
title = {New TPC benchmarks for decision support and web commerce},
year = {2000},
issue_date = {Dec. 2000},
publisher = {Association for Computing Machinery},
address = {New York, NY, USA},
volume = {29},
number = {4},
issn = {0163-5808},
url = {https://doi.org/10.1145/369275.369291},
doi = {10.1145/369275.369291},
journal = {SIGMOD Rec.},
month = dec,
pages = {64–71},
numpages = {8}
}

@article{rbot,
  author       = {Zhaoyan Sun and
                  Xuanhe Zhou and
                  Guoliang Li and
                  Xiang Yu and
                  Jianhua Feng and
                  Yong Zhang},
  title        = {R-Bot: An LLM-based Query Rewrite System},
  journal      = {Proc. {VLDB} Endow.},
  volume       = {18},
  number       = {12},
  pages        = {5031--5044},
  year         = {2025},
  url          = {https://www.vldb.org/pvldb/vol18/p5031-li.pdf},
  doi          = {10.14778/3750601.3750625},
  bibsource    = {dblp computer science bibliography, https://dblp.org}
}

@article{gpt-4o,
  author       = {OpenAI},
  title        = {GPT-4o System Card},
  journal      = {CoRR},
  volume       = {abs/2410.21276},
  year         = {2024},
  url          = {https://doi.org/10.48550/arXiv.2410.21276},
  doi          = {10.48550/ARXIV.2410.21276},
  eprinttype   = {arXiv},
  eprint       = {2410.21276},
  bibsource    = {dblp computer science bibliography, https://dblp.org}
}

@article{gemini-2.5,
  author       = {Gemini Team},
  title        = {Gemini 2.5: Pushing the Frontier with Advanced Reasoning, Multimodality,
                  Long Context, and Next Generation Agentic Capabilities},
  journal      = {CoRR},
  volume       = {abs/2507.06261},
  year         = {2025},
  url          = {https://doi.org/10.48550/arXiv.2507.06261},
  doi          = {10.48550/ARXIV.2507.06261},
  eprinttype   = {arXiv},
  eprint       = {2507.06261},
  bibsource    = {dblp computer science bibliography, https://dblp.org}
}

@misc{openai_gpt54_thinking,
  title        = {{GPT-5.4 Thinking System Card}},
  author       = {{OpenAI}},
  year         = {2026},
  month        = mar,
  howpublished = {\url{https://deploymentsafety.openai.com/gpt-5-4-thinking/gpt-5-4-thinking.pdf}},
}

@article{qwen3,
  author       = {Qwen Team},
  title        = {Qwen3 Technical Report},
  journal      = {CoRR},
  volume       = {abs/2505.09388},
  year         = {2025},
  url          = {https://doi.org/10.48550/arXiv.2505.09388},
  doi          = {10.48550/ARXIV.2505.09388},
  eprinttype   = {arXiv},
  eprint       = {2505.09388},
  bibsource    = {dblp computer science bibliography, https://dblp.org}
}

@phdthesis{DBLP:phd/Graefe87,
  author       = {Goetz Graefe},
  title        = {Rule-Based Query Optimization in Extensible Database Systems},
  school       = {Univ. of Wisconsin-Madison},
  year         = {1987},
  bibsource    = {dblp computer science bibliography, https://dblp.org}
}

@inproceedings{DBLP:conf/vldb/LevyMS94,
  author       = {Alon Y. Levy and
                  Inderpal Singh Mumick and
                  Yehoshua Sagiv},
  editor       = {Jorge B. Bocca and
                  Matthias Jarke and
                  Carlo Zaniolo},
  title        = {Query Optimization by Predicate Move-Around},
  booktitle    = {VLDB'94, Proceedings of 20th International Conference on Very Large
                  Data Bases, September 12-15, 1994, Santiago de Chile, Chile},
  pages        = {96--107},
  publisher    = {Morgan Kaufmann},
  year         = {1994},
  url          = {http://www.vldb.org/conf/1994/P096.PDF},
  bibsource    = {dblp computer science bibliography, https://dblp.org}
}

@inproceedings{DBLP:conf/vldb/Muralikrishna92,
  author       = {M. Muralikrishna},
  editor       = {Li{-}Yan Yuan},
  title        = {Improved Unnesting Algorithms for Join Aggregate {SQL} Queries},
  booktitle    = {18th International Conference on Very Large Data Bases, August 23-27,
                  1992, Vancouver, Canada, Proceedings},
  pages        = {91--102},
  publisher    = {Morgan Kaufmann},
  year         = {1992},
  url          = {http://www.vldb.org/conf/1992/P091.PDF},
  bibsource    = {dblp computer science bibliography, https://dblp.org}
}

@article{LeisGMBK015,
  author       = {Viktor Leis and
                  Andrey Gubichev and
                  Atanas Mirchev and
                  Peter Boncz and
                  Alfons Kemper and
                  Thomas Neumann},
  title        = {How Good Are Query Optimizers, Really?},
  journal      = {Proc. {VLDB} Endow.},
  volume       = {9},
  number       = {3},
  pages        = {204--215},
  year         = {2015},
  url          = {http://www.vldb.org/pvldb/vol9/p204-leis.pdf},
  doi          = {10.14778/2850583.2850594},
  bibsource    = {dblp computer science bibliography, https://dblp.org}
}

@article{LeisGMBKN25,
  author       = {Viktor Leis and
                  Andrey Gubichev and
                  Atanas Mirchev and
                  Peter Boncz and
                  Alfons Kemper and
                  Thomas Neumann},
  title        = {Still Asking: How Good Are Query Optimizers, Really?},
  journal      = {Proc. {VLDB} Endow.},
  volume       = {18},
  number       = {12},
  pages        = {5531--5536},
  year         = {2025},
  url          = {https://www.vldb.org/pvldb/vol18/p5531-viktor.pdf},
  doi          = {10.14778/3750601.3760521},
  bibsource    = {dblp computer science bibliography, https://dblp.org}
}

@article{slabcity,
  author       = {Rui Dong and
                  Jie Liu and
                  Yuxuan Zhu and
                  Cong Yan and
                  Barzan Mozafari and
                  Xinyu Wang},
  title        = {SlabCity: Whole-Query Optimization using Program Synthesis},
  journal      = {Proc. {VLDB} Endow.},
  volume       = {16},
  number       = {11},
  pages        = {3151--3164},
  year         = {2023},
  url          = {https://www.vldb.org/pvldb/vol16/p3151-dong.pdf},
  doi          = {10.14778/3611479.3611515},
  bibsource    = {dblp computer science bibliography, https://dblp.org}
}

@article{grpo,
  author       = {Zhihong Shao and
                  Peiyi Wang and
                  Qihao Zhu and
                  Runxin Xu and
                  Junxiao Song and
                  Mingchuan Zhang and
                  Y. K. Li and
                  Y. Wu and
                  Daya Guo},
  title        = {DeepSeekMath: Pushing the Limits of Mathematical Reasoning in Open
                  Language Models},
  journal      = {CoRR},
  volume       = {abs/2402.03300},
  year         = {2024},
  url          = {https://doi.org/10.48550/arXiv.2402.03300},
  doi          = {10.48550/ARXIV.2402.03300},
  eprinttype   = {arXiv},
  eprint       = {2402.03300},
  bibsource    = {dblp computer science bibliography, https://dblp.org}
}

@article{sefrqo,
  author       = {Hanwen Liu and
                  Qihan Zhang and
                  Ryan Marcus and
                  Ibrahim Sabek},
  title        = {{SEFRQO:} {A} Self-Evolving Fine-Tuned RAG-Based Query Optimizer},
  journal      = {Proc. {ACM} Manag. Data},
  volume       = {3},
  number       = {6},
  pages        = {1--27},
  year         = {2025},
  url          = {https://doi.org/10.1145/3769826},
  doi          = {10.1145/3769826},
  bibsource    = {dblp computer science bibliography, https://dblp.org}
}

@article{e3rewrite,
  author       = {Dongjie Xu and
                  Yue Cui and
                  Weijie Shi and
                  Qingzhi Ma and
                  Hanghui Guo and
                  Jiaming Li and
                  Yao Zhao and
                  Ruiyuan Zhang and
                  Shimin Di and
                  Jia Zhu and
                  Kai Zheng and
                  Jiajie Xu},
  title        = {E3-Rewrite: Learning to Rewrite {SQL} for Executability, Equivalence,and
                  Efficiency},
  journal      = {CoRR},
  volume       = {abs/2508.09023},
  year         = {2025},
  url          = {https://doi.org/10.48550/arXiv.2508.09023},
  doi          = {10.48550/ARXIV.2508.09023},
  eprinttype   = {arXiv},
  eprint       = {2508.09023},
  bibsource    = {dblp computer science bibliography, https://dblp.org}
}

@inproceedings{calcite,
  author       = {Edmon Begoli and
                  Jes{\'{u}}s Camacho{-}Rodr{\'{\i}}guez and
                  Julian Hyde and
                  Michael J. Mior and
                  Daniel Lemire},
  editor       = {Gautam Das and
                  Christopher M. Jermaine and
                  Philip A. Bernstein},
  title        = {Apache Calcite: {A} Foundational Framework for Optimized Query Processing
                  Over Heterogeneous Data Sources},
  booktitle    = {Proceedings of the 2018 International Conference on Management of
                  Data, {SIGMOD} Conference 2018, Houston, TX, USA, June 10-15, 2018},
  pages        = {221--230},
  publisher    = {{ACM}},
  year         = {2018},
  url          = {https://doi.org/10.1145/3183713.3190662},
  doi          = {10.1145/3183713.3190662},
  bibsource    = {dblp computer science bibliography, https://dblp.org}
}

@inproceedings{overview,
author = {Chaudhuri, Surajit},
title = {An overview of query optimization in relational systems},
year = {1998},
isbn = {0897919963},
publisher = {Association for Computing Machinery},
address = {New York, NY, USA},
url = {https://doi.org/10.1145/275487.275492},
doi = {10.1145/275487.275492},
booktitle = {Proceedings of the Seventeenth ACM SIGACT-SIGMOD-SIGART Symposium on Principles of Database Systems},
pages = {34–43},
numpages = {10},
location = {Seattle, Washington, USA},
series = {PODS '98}
}

@article{path_selection,
  author       = {Laura M. Haas},
  title        = {Review - Access Path Selection in a Relational Database Management
                  System},
  journal      = {{ACM} {SIGMOD} Digit. Rev.},
  volume       = {1},
  year         = {1999},
  url          = {https://dblp.org/db/journals/dr/Haas99a.html},
  bibsource    = {dblp computer science bibliography, https://dblp.org}
}

@article{DBLP:journals/ftdb/DingNC24,
  author       = {Bailu Ding and
                  Vivek R. Narasayya and
                  Surajit Chaudhuri},
  title        = {Extensible Query Optimizers in Practice},
  journal      = {Found. Trends Databases},
  volume       = {14},
  number       = {3-4},
  pages        = {186--402},
  year         = {2024},
  url          = {https://doi.org/10.1561/1900000077},
  doi          = {10.1561/1900000077},
  bibsource    = {dblp computer science bibliography, https://dblp.org}
}

@article{DBLP:journals/corr/abs-2502-12918,
  author       = {Sriram Dharwada and
                  Himanshu Devrani and
                  Jayant R. Haritsa and
                  Harish Doraiswamy},
  title        = {Query Rewriting via LLMs},
  journal      = {CoRR},
  volume       = {abs/2502.12918},
  year         = {2025},
  url          = {https://doi.org/10.48550/arXiv.2502.12918},
  doi          = {10.48550/ARXIV.2502.12918},
  eprinttype   = {arXiv},
  eprint       = {2502.12918},
  bibsource    = {dblp computer science bibliography, https://dblp.org}
}

@article{DBLP:journals/tods/TrummerWWMMJAR21,
  author       = {Immanuel Trummer and
                  Junxiong Wang and
                  Ziyun Wei and
                  Deepak Maram and
                  Samuel Moseley and
                  Saehan Jo and
                  Joseph Antonakakis and
                  Ankush Rayabhari},
  title        = {SkinnerDB: Regret-bounded Query Evaluation via Reinforcement Learning},
  journal      = {{ACM} Trans. Database Syst.},
  volume       = {46},
  number       = {3},
  pages        = {9:1--9:45},
  year         = {2021},
  url          = {https://doi.org/10.1145/3464389},
  doi          = {10.1145/3464389},
  bibsource    = {dblp computer science bibliography, https://dblp.org}
}

@article{neo,
  author       = {Ryan Marcus and
                  Parimarjan Negi and
                  Hongzi Mao and
                  Chi Zhang and
                  Mohammad Alizadeh and
                  Tim Kraska and
                  Olga Papaemmanouil and
                  Nesime Tatbul},
  title        = {Neo: {A} Learned Query Optimizer},
  journal      = {Proc. {VLDB} Endow.},
  volume       = {12},
  number       = {11},
  pages        = {1705--1718},
  year         = {2019},
  url          = {http://www.vldb.org/pvldb/vol12/p1705-marcus.pdf},
  doi          = {10.14778/3342263.3342644},
  bibsource    = {dblp computer science bibliography, https://dblp.org}
}

@inproceedings{LSTM,
  author       = {Xiang Yu and
                  Guoliang Li and
                  Chengliang Chai and
                  Nan Tang},
  title        = {Reinforcement Learning with Tree-LSTM for Join Order Selection},
  booktitle    = {36th {IEEE} International Conference on Data Engineering, {ICDE} 2020,
                  Dallas, TX, USA, April 20-24, 2020},
  pages        = {1297--1308},
  publisher    = {{IEEE}},
  year         = {2020},
  url          = {https://doi.org/10.1109/ICDE48307.2020.00116},
  doi          = {10.1109/ICDE48307.2020.00116},
  bibsource    = {dblp computer science bibliography, https://dblp.org}
}

@inproceedings{Balsa,
  author       = {Zongheng Yang and
                  Wei{-}Lin Chiang and
                  Sifei Luan and
                  Gautam Mittal and
                  Michael Luo and
                  Ion Stoica},
  editor       = {Zachary G. Ives and
                  Angela Bonifati and
                  Amr El Abbadi},
  title        = {Balsa: Learning a Query Optimizer Without Expert Demonstrations},
  booktitle    = {{SIGMOD} '22: International Conference on Management of Data, Philadelphia,
                  PA, USA, June 12 - 17, 2022},
  pages        = {931--944},
  publisher    = {{ACM}},
  year         = {2022},
  url          = {https://doi.org/10.1145/3514221.3517885},
  doi          = {10.1145/3514221.3517885},
  bibsource    = {dblp computer science bibliography, https://dblp.org}
}

@inproceedings{dba,
  author       = {R. Malinga Perera and
                  Bastian Oetomo and
                  Benjamin I. P. Rubinstein and
                  Renata Borovica{-}Gajic},
  title        = {{DBA} bandits: Self-driving index tuning under ad-hoc, analytical
                  workloads with safety guarantees},
  booktitle    = {37th {IEEE} International Conference on Data Engineering, {ICDE} 2021,
                  Chania, Greece, April 19-22, 2021},
  pages        = {600--611},
  publisher    = {{IEEE}},
  year         = {2021},
  url          = {https://doi.org/10.1109/ICDE51399.2021.00058},
  doi          = {10.1109/ICDE51399.2021.00058},
  bibsource    = {dblp computer science bibliography, https://dblp.org}
}

@article{udo,
  author       = {Junxiong Wang and
                  Immanuel Trummer and
                  Debabrota Basu},
  title        = {{UDO:} Universal Database Optimization using Reinforcement Learning},
  journal      = {Proc. {VLDB} Endow.},
  volume       = {14},
  number       = {13},
  pages        = {3402--3414},
  year         = {2021},
  url          = {http://www.vldb.org/pvldb/vol14/p3402-wang.pdf},
  doi          = {10.14778/3484224.3484236},
  bibsource    = {dblp computer science bibliography, https://dblp.org}
}

@article{DBLP:journals/pvldb/GaoWLSQDZ24,
  author       = {Dawei Gao and
                  Haibin Wang and
                  Yaliang Li and
                  Xiuyu Sun and
                  Yichen Qian and
                  Bolin Ding and
                  Jingren Zhou},
  title        = {Text-to-SQL Empowered by Large Language Models: {A} Benchmark Evaluation},
  journal      = {Proc. {VLDB} Endow.},
  volume       = {17},
  number       = {5},
  pages        = {1132--1145},
  year         = {2024},
  url          = {https://www.vldb.org/pvldb/vol17/p1132-gao.pdf},
  doi          = {10.14778/3641204.3641221},
  bibsource    = {dblp computer science bibliography, https://dblp.org}
}

@article{SQL-Factory,
  author       = {Jiahui Li and
                  Tongwang Wu and
                  Yuren Mao and
                  Yunjun Gao and
                  Yajie Feng and
                  Huaizhong Liu},
  title        = {SQL-Factory: {A} Multi-Agent Framework for High-Quality and Large-Scale
                  {SQL} Generation},
  journal      = {Proc. {VLDB} Endow.},
  volume       = {19},
  number       = {3},
  pages        = {292--305},
  year         = {2025},
  url          = {https://www.vldb.org/pvldb/vol19/p292-gao.pdf},
  bibsource    = {dblp computer science bibliography, https://dblp.org}
}

@article{OpenSearch-SQL,
  author       = {Xiangjin Xie and
                  Guangwei Xu and
                  Lingyan Zhao and
                  Ruijie Guo},
  title        = {OpenSearch-SQL: Enhancing Text-to-SQL with Dynamic Few-shot and Consistency
                  Alignment},
  journal      = {Proc. {ACM} Manag. Data},
  volume       = {3},
  number       = {3},
  pages        = {194:1--194:24},
  year         = {2025},
  url          = {https://doi.org/10.1145/3725331},
  doi          = {10.1145/3725331},
  bibsource    = {dblp computer science bibliography, https://dblp.org}
}

@article{DBLP:journals/pacmmod/ChenCKY25,
  author       = {Kaiwen Chen and
                  Yueting Chen and
                  Nick Koudas and
                  Xiaohui Yu},
  title        = {Reliable Text-to-SQL with Adaptive Abstention},
  journal      = {Proc. {ACM} Manag. Data},
  volume       = {3},
  number       = {1},
  pages        = {69:1--69:30},
  year         = {2025},
  url          = {https://doi.org/10.1145/3709719},
  doi          = {10.1145/3709719},
  bibsource    = {dblp computer science bibliography, https://dblp.org}
}

@article{DBLP:journals/pacmmod/YangWXWDPCL25,
  author       = {Yicun Yang and
                  Zhaoguo Wang and
                  Yu Xia and
                  Zhuoran Wei and
                  Haoran Ding and
                  Ruzica Piskac and
                  Haibo Chen and
                  Jinyang Li},
  title        = {Automated Validating and Fixing of Text-to-SQL Translation with Execution Consistency},
  journal      = {Proc. {ACM} Manag. Data},
  volume       = {3},
  number       = {3},
  pages        = {134:1--134:28},
  year         = {2025},
  url          = {https://doi.org/10.1145/3725271},
  doi          = {10.1145/3725271},
  bibsource    = {dblp computer science bibliography, https://dblp.org}
}
